\documentclass[12pt]{iopart}

\usepackage{textcomp,graphicx,bm,xcolor}
\usepackage{amsmath}
\begin{document}

\title[PSub flow stabilization by elastic metamaterials]{Phononic-subsurface flow stabilization by subwavelength locally resonant metamaterials}

\author{Armin Kianfar}
\address{Ann and H.J. Smead Department of Aerospace Engineering Sciences, University of Colorado Boulder, Boulder, Colorado 80303, USA}
\author{Mahmoud I. Hussein\footnote{Corresponding author: Mahmoud I. Hussein (mih@colorado.edu)}}

\address{Ann and H.J. Smead Department of Aerospace Engineering Sciences, University of Colorado Boulder, Boulder, Colorado 80303, USA}
\address{Department of Physics, University of Colorado Boulder, Boulder, Colorado 80302, USA}

%\ead{mih@colorado.edu}
\vspace{10pt}
%\begin{indented}
%\date{\today} 
%\end{indented}

\begin{abstract}
The interactions between a solid surface and a fluid flow underlie dynamical processes relevant to air, sea, and land vehicle performance and numerous other technologies. Key among these processes are unstable flow disturbances that contribute to fundamental transformations in the flow field.~Precise control of these disturbances is possible by introducing a \textit{phononic subsurface} (PSub).~This comprises locally attaching a finite phononic structure perpendicular to an elastic surface exposed to the flowing fluid. This structure experiences ongoing excitation by an unstable flow mode, or more than one mode, traveling in conjunction with the mean flow. The excitation generates small deformations at the surface that trigger elastic wave propagation within the structure, traveling away from the flow and reflecting at the end of the structure to return to the fluid-structure interface and back into the flow. By targeted tuning of the unit-cell and finite-structure characteristics of the PSub, the returning waves may be devised to resonate and reenter the flow out of phase, leading to significant destructive interference of the continuously incoming flow waves near the surface and subsequently to their attenuation over the spatial extent of the control region. This entire mechanism is passive, responsive, and engineered offline without needing coupled fluid-structure simulations; only the flow instability's frequency, wavelength, and overall modal characteristics must be known.~Disturbance stabilization in a wall-bounded transitional flow leads to delay in laminar-to-turbulent transition and reduction in skin-friction drag.~Destabilization is also possible by alternatively designing the PSub to induce constructive interference, which is beneficial for delaying flow separation and enhancing chemical mixing and combustion. In this paper, we present a PSub in the form of a locally resonant elastic metamaterial, designed to operate in the elastic subwavelength regime and hence being significantly shorter in length compared to a phononic-crystal-based PSub.~This is enabled by utilizing a sub-hybridization resonance.~Using direct numerical simulations (DNS) of channel flows, both types of PSubs are investigated, and their controlled spatial and energetic influence on the wall-bounded flow behavior is demonstrated and analyzed.~We show that the PSub's effect is spatially localized as intended, with a rapidly diminishing streamwise influence away from its location in the subsurface.
\end{abstract}

%\maketitle

\section{Introduction}
\label{Sec:Introduction}
Flow control is a central topic in fluid dynamics that is concerned with devising passive or active means of intervention with the flow structure and its underlying mechanisms in a manner that causes desirable changes in the overall flow behavior. Through flow control, it is possible, in principle, to enable favorable outcomes such as, for example, delay of laminar-to-turbulent transition and reduction of skin-friction drag in wall-bounded flows~\cite{Gad-el-Hak_2000}. These scenarios allow for substantial savings in fuel expenditure for air, sea, land vehicles, wind and water turbines, long-range gas and liquid pipelines, and other similar applications. Flow control by active means has been extensively investigated over the past few decades~\cite{Wehrmann_1965,Liepmann_1982,Joslin_1995,Grundmann_2008,Amitay_2016,Jansen_2018}. Passive techniques, on the other hand, are desirable because of their simplicity and low cost, i.e., no active control devices, wires, ducts, slots, etc., are needed and no electric power is required to drive the control process. Passive techniques widely explored in the literature include the use of riblets~\cite{Walsh_1978,Garcia_2011}, roughness~\cite{Cossu_2002,Fransson_2005}, or porous features~\cite{Abderrahaman_2017} on the surface exposed to the flow, or coating the surface with a compliant material~\cite{Kramer_1957,Benjamin_1960,Bushnell_1977,Gad-el-Hak_1984,Carpenter_1985,Lucy_1995,Davies_1997,Luhar_2015,Esteghamatian_2022}. An ideal intervention requires an understanding of the key characteristics of the flow dynamics and using this knowledge to tailor, with dynamical precision, a control stimulus that accounts for the underlying flow mechanisms. Recently, this endeavor has been shown to be possible with the use of phononic materials passively employed in the subsurface of a wall-bounded flow~\cite{Hussein_2015,Barnes_2021}.

\begin{figure*} 
	\includegraphics{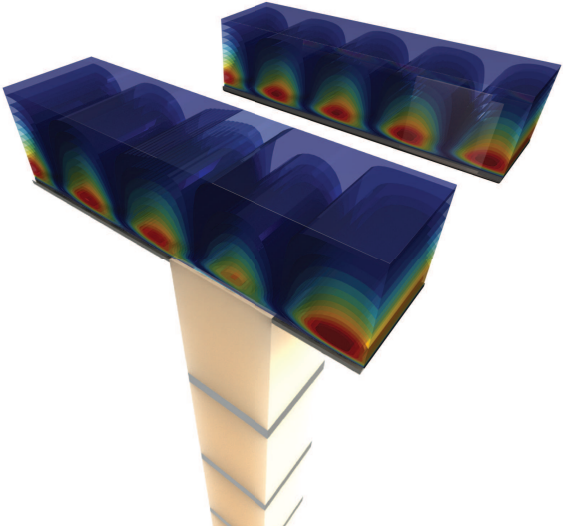}%
	\caption{Passive flow stabilization by a PSub.~Contours showing the streamwise-component of an instability velocity field when a PnC-based PSub is installed (front) versus an all-rigid-wall surface (back)~\cite{Hussein_2015}. Yellow color represents low-instability intensity, red color represents high-instability intensity. The reduced color intensity at the position where the PSub is placed is indicative of stabilization exactly at that location. } \label{Fig1}
\end{figure*}

Flow transition may occur when external disturbances or inherent fluctuations develop and become significant within the flow field. These disturbances may be in the form of unstable waves that represent a small component of the total velocity field; an example of a widely studied type of disturbance in shear flows is a Tollmien-Schlichting (TS) wave~\cite{Tollmien_1928,Schlichting_1933}. In this context, flow disturbances, also known as perturbations or instabilities\footnote{We will use the terms \textit{perturbation} and \textit{instability}, interchangeably, when referring to the flow waves.}, appear at various frequencies, wavenumbers, phases, and orientations and depending on their character may grow in amplitude as they travel downstream. In 2015, the general concept of a \textit{phononic subsurface} (PSub)~\cite{Hussein_2015} was introduced as a means to provide a wave-synchronized intervention with flow instabilities to cause either stabilization or destabilization, as desired.~A PSub is installed in the subsurface region, and is nominally perpendicularly oriented and configured to extend all the way to expose its edge to the flow, forming an elastic fluid-structure interface.~The underlying mechanism that a PSub induces is passive and responsive\footnote{Responsive control implies that no phase locking is required. The PSub adaptively responds as desired regardless of the specific phase of the incoming instability wave when it arrives at the control region.} localized control of both the sign and rate of production of the perturbation kinetic energy within the flow field.~A strong (weak) PSub intervention for flow stabilization causes a strong (weak) negative rate of production, effectively shutting off the source of energy intake into the instability from the mean flow.~When a PSub is introduced for flow destabilization, the opposite effect takes place and the instability is forced to acquire energy from the mean flow at a higher rate.~These two scenarios are manifestations of a contiguous solid-fluid flow antiresonance or resonance phenomenon, respectively.~A PSub takes the form of a finite, relatively stiff elastic structure oriented in a manner that enables only small elastic motion perpendicular to the fluid-structure interface to be admitted and transferred into the flow.~Ensuring small vibrations at the surface allows the PSub to modulate mostly (to the extent allowed in practice) a single-velocity component of the instability field (the wall-normal component in the present study), as opposed to simultaneously influencing multiple components at once.~With these conditions in place, the PSub is engineered to exhibit specific frequency-dependent \textit{amplitude} and \textit{phase} response characteristics at the edge exposed to the flow.~These quantities represent the two core properties on which the PSubs design theory is based on.~Figure~\ref{Fig1} provides an illustration of a PSub in operation, showing clearly its ability to attenuate the instability field exactly within the region in the flow where the PSub is installed~\cite{Hussein_2015}.

\begin{figure*} [h!]
	\includegraphics{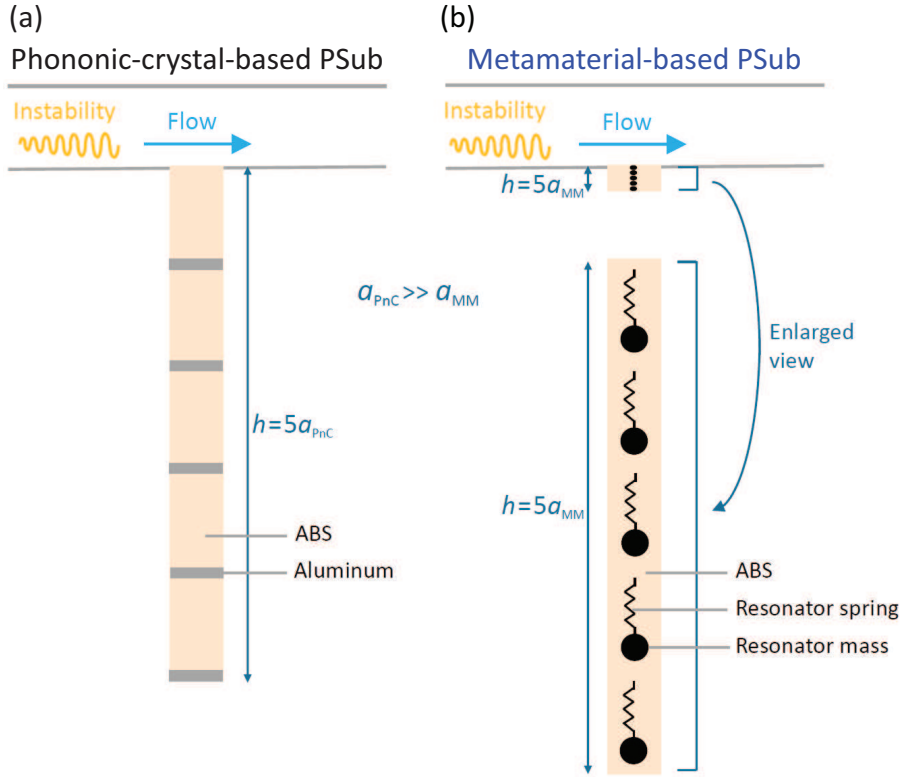}%
	\caption{Schematic of two types of PSubs: (a) a phononic-crystal-based PSub~\cite{Hussein_2015} and (b) a locally-resonant elastic metamaterial-based PSub. In this schematic, the PSub in (b) has a unit-cell length 40 times shorter than the PSub in (a). Each PSub is installed in the flow subsurface and extends all the way to allow for direct exposition to the flow. Flow instabilities, e.g., TS waves, will excite the PSub at the top edge (i.e., at the fluid-structure interface), and the PSub, in turn, will respond at or near its structural resonance and out of phase at the excitation point. This passive process will repeat and cause sustained attenuation of reoccurring and continuously incoming instability waves. Alternatively, the PSub could be designed to trigger destabilization instead of stabilization by producing an in-phase elastic response.} \label{Fig2}
\end{figure*}

A PSub may be composed of any form of phononic materials\footnote{From a configurational perspective, a PSub, in general, can take any form that achieves the mechanistic intervention with the flow production rate described above.~For example a standard homogeneous and uniform finite structure may be employed. However, a phononic structure provides significantly favorable dynamical properties and attributes because a phononic material has a frequency band structure~\cite{Hussein_2014,Jin_2021} and when rendered finite gives unique local and global resonance characteristics~\cite{wallis1957effect,Camley_1983,Davis_2011,Albabaa_2017,Bastawrous_2022,Albabaa_2022,Rosa_2022} that are not attainable by conventional materials and that are highly controllable by design.}.~The study of phononic materials, in general, is an area that has received tremendous attention in the literature over the past three decades~\cite{Hussein_2014,Jin_2021}. Figure~\ref{Fig2} displays a schematic of two types of PSubs. The model in Fig.~\ref{Fig2}a is based on a phononic-crystal (PnC) rod comprised of a repeated layering of acrylonitrile butadiene styrene (ABS) polymer and aluminum; this is the original configuration used in Ref.~\cite{Hussein_2015}. In contrast, the PSub configuration in Fig.~\ref{Fig2}b is formed from a locally-resonant elastic metamaterial (MM) which is here realized in the form of a homogeneous ABS polymer rod with a repeated inclusion of spring-mass units to serve as intrinsic local resonators.~Both structures comprise five unit cells in the schematic and throughout the paper. Phononic crystals draw their unique wave propagation properties from wave interference mechanisms, namely Bragg scattering~\cite{Kushwaha_1993}.~This typically requires the unit cells to be relatively long for intervention at a given frequency regime; e.g., the unit cell used in Ref.~\cite{Hussein_2015} is 40 cm long to enable control of a flow instability near 2~kHz.~A 10-unit-cell PSub, in that case, would be 4~m long extending into the subsurface in the wall-normal direction, prohibiting practical deployment.~The schematic shown in Fig.~\ref{Fig1} is of this particular PSub, passively stabilizing a TS wave~\cite{Hussein_2015}.~Elastic metamaterials, on the other hand, produce their unique wave propagation properties via resonance hybridization$-$an elastodynamic coupling mechanism that frees the unit cell from any length constraints~\cite{Liu_2000}.~An alternative PSub configuration for overcoming this length constraint is a coiled PnC~\cite{Barnes_2021,Willey_2022}. The reader may refer to books~\cite{Deymier_2013,Craster_2013,Phani_2017} and extensive reviews~\cite{Hussein_2014,Jin_2021} for in-depth description and analysis of phononic crystals and elastic metamaterials.~Another key PSub dimension is its length along the streamwise direction; this has to be designed to correlate with the wavelength of the instability waves (or range of wavelengths in the case of multiple instability waves).   

In this paper, the notion of a metamaterial-based PSub is introduced to mitigate the large unit-cell length limitation imposed on a phononic-crystal-based PSub.~Furthermore, we demonstrate the desirable precision of the PSub's impact on the flow field, showing that the alterations to the flow structure are spatially localized, as targeted, with no (or insignificant) undesirable behavior downstream to the position in the flow where a PSub is installed.~The type and intensity of control are also shown to take effect with design precision.~Finally, we provide a rigorous analysis of the effect of the PSub on the intrinsic flow dynamics, quantitatively demonstrating the mechanism of the rate of energy exchange between a flow instability and the mean flow as a result of the presence of a PSub.~We also examine the PSub's spatial influence on the flow vector field, and, conversely, the flow instability’s spatial influence on the elastodynamic energy field within the PSub itself. The unique ability to passively enact perfect wave synchronization across the PSub and flow domains is explicitly demonstrated.

\begin{figure*} [t!]
	\includegraphics[width=0.6\columnwidth]{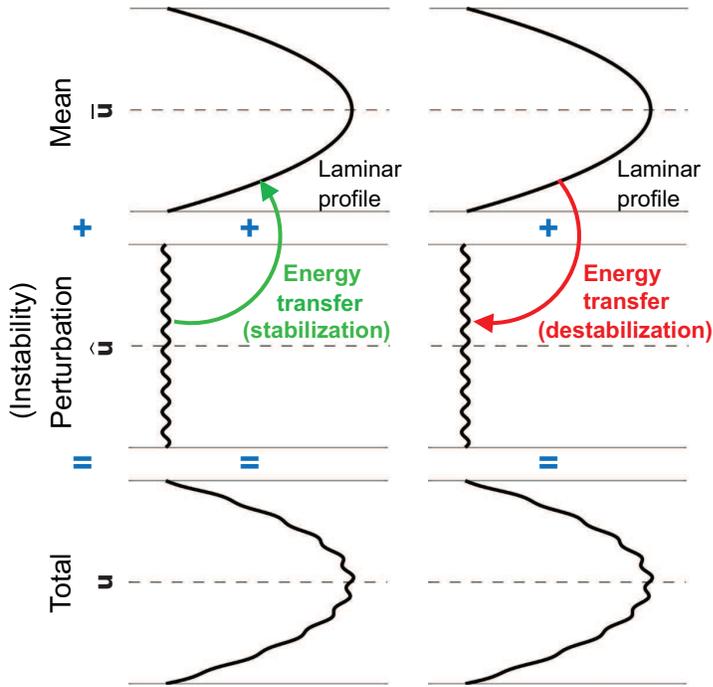}%
	\caption{Schematic illustration of key energy exchange mechanisms passively triggered by a PSub installed for channel flow control.~The flow total velocity field $\mathbf{u}$ is decomposed into a mean flow component $\bar{\mathbf{u}}$ and a perturbation (instability) component $\hat{\mathbf{u}}$. When designed for stabilization, the PSub causes energy transfer from the instability component to the mean flow component. When designed for destabilization, the opposite effect takes place.~The flow model details are given in Section~\ref{FlowModel}} \label{Fig3}
\end{figure*}

\section{PSub Design Theory}
\label{sec:PSubs}
The general elements of the PSub design theory were outlined in Ref.~\cite{Hussein_2015}.~A PSub configuration consists of a finite phononic structure with its principle path of elastic wave propagation typically oriented orthogonal to the fluid-structure interface to enable ``pointwise" spatial control as needed. A PSub is engineered offline to exhibit target frequency-dependent amplitude and phase response characteristics at the edge exposed to the flow, i.e., at the top end of each PSub shown in Fig.~\ref{Fig2}. This pair of response quantities at this location represents the two principle properties targeted by the PSub design theory. In all cases, the PSub edge should be ensured to vibrate at, or close to, resonance at the frequency of the instability to be controlled. A high vibration amplitude allows for strong interaction with the flow. Yet, still, the regime of operation is intentionally limited to small elastic vibrations, where the local fluid-structure interface remains practically flat, and large finite deformations of the solid surface are avoided.~As described above, this confines the control to exclusively, or predominantly, the vertical, i.e., wall-normal, component of the perturbation velocity field (see Section~\ref{sec:Res} for an analysis and further discussion on this aspect).~As for the phase, the PSub is designed to display a negative phase (out of phase) if the target is stabilization or a positive phase (in phase) if the target is destabilization.~Given the importance of both the vibration amplitude and phase, a performance metric $P$ was introduced and is defined as the frequency-dependent product of the two quantities. Negative and positive values of $P$ correspond to flow stabilization and destabilization, respectively. The absolute value $|P|$ indicates the strength of the stabilization or destabilization. For example, to impede the growth of a particular instability to delay the transition to turbulence, the PSub is designed to exhibit a strongly negative $P$ value at the frequency of the instability. For a range of instability frequencies, the PSub would need to display this property over that frequency range.~As for the spatial size or width of a PSub along the downstream direction; this is tuned according to the wavelength of the flow wave instability to be controlled.~In Ref.~\cite{Hussein_2015}, the PSub length was set to be roughly one quarter of the wavelength of the unstable flow wave. 

From the flow's perspective, the phase of the elastic waves returning to the flow$-$after being passively processed by the PSub$-$will cause destructive or constructive interference with the vertical velocity component of the continuously incoming instability waves. This, in turn, will influence the work done by or on the instability field, causing either a diminishing or an enhancing effect on the transfer of energy from the mean flow into the instability, depending on whether the PSub is designed to stabilize or destabilize, respectively.~Figure~\ref{Fig3} provides a schematic illustration of this mechanism.~This effect on the exchange of energy with the mean flow is quantified by what is known as the production rate term, an averaged quantity involving the wall-normal and streamwise (vertical and horizontal, respectively, in Fig.~\ref{Fig2}) components of the instability field, which is derived from the Navier-Stokes equations governing the flow~\cite{Hussein_2015}.
\begin{figure*} [b!]
	\includegraphics[width=1\columnwidth]{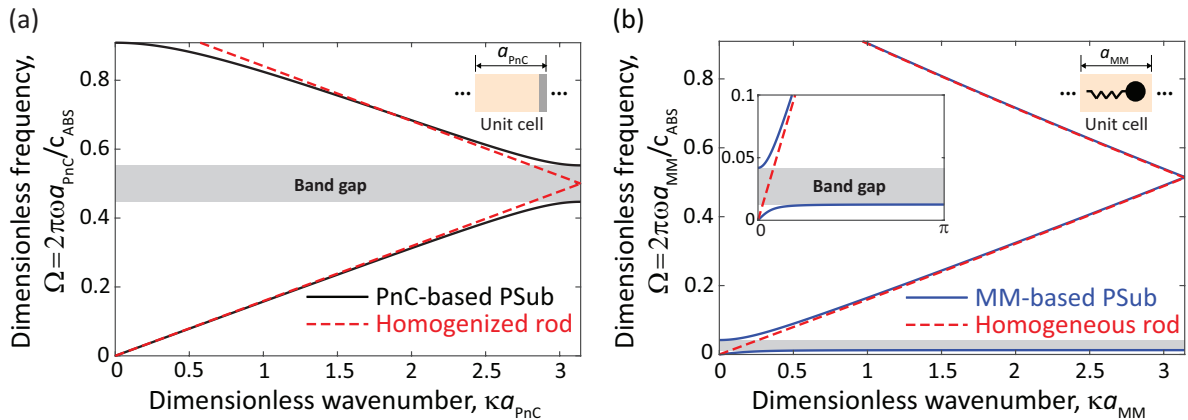}%
	\caption{Dispersion diagram for (a) PnC unit cell and (b) locally resonant elastic MM unit cell with a close-up.~The dispersion curves for a corresponding homogeneous rod unit cell in each case are also provided.~Schematics of the unit-cell configurations are shown as insets.~The frequency and wavenumber are nondimensionalized by multiplication with corresponding unit-cell parameters: $a_{\rm PnC}$ and $a_{\rm MM}$ are the PnC and MM unit-cell length, respectively, and $c_{\rm ABS}$ is the long-wave longitudinal speed for the ABS material.} \label{Fig4}
\end{figure*}
\subsection{Stop-band truncation-resonance approach}
\label{sec:PSubsStop}
\begin{figure*} [b!]
	\includegraphics[width=1\columnwidth]{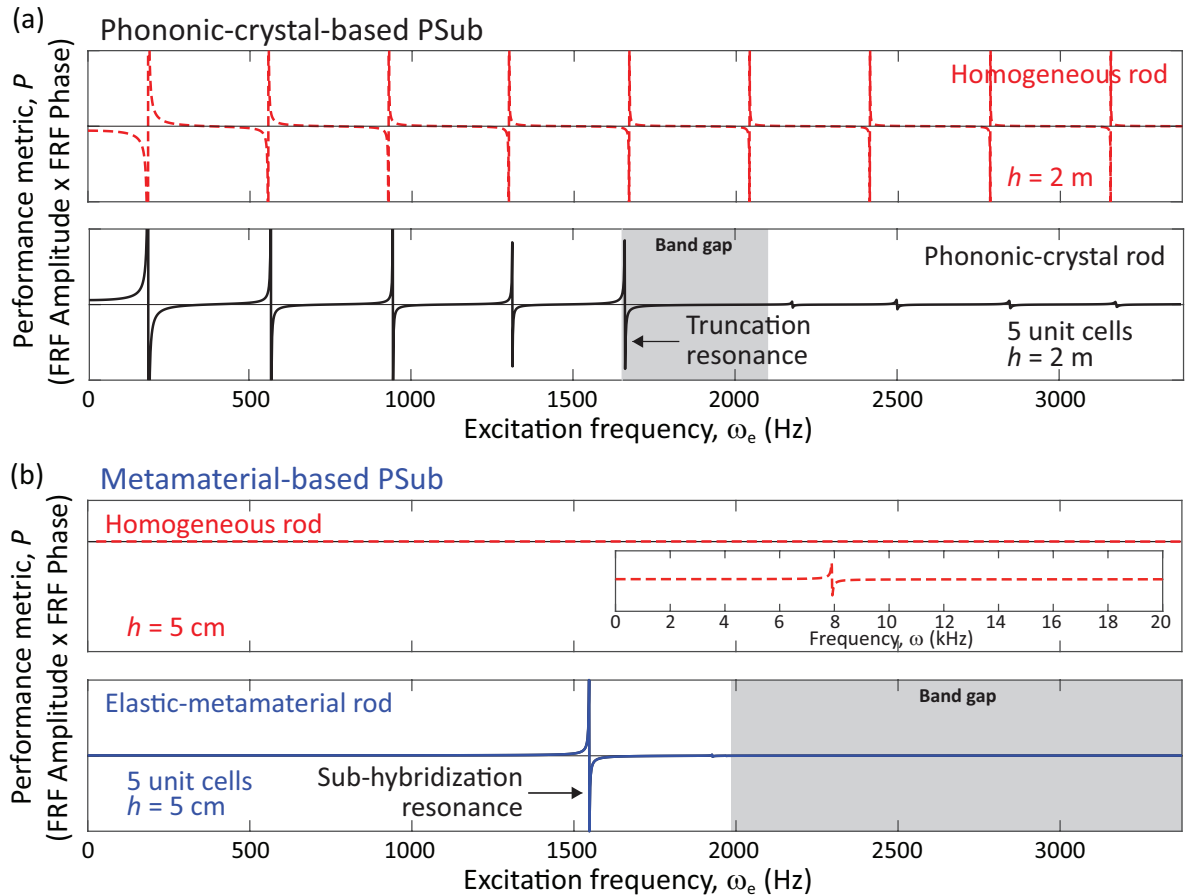}%
	\caption{Schematic illustration of two PSub design principles: PnC-based PSub versus MM-based PSub. To produce the desired performance metric properties at the frequency of the instability, in (a) a truncation resonance is utilized (PnC-based PSub), and in (b) a sub-hybridization resonance is utilized (MM-based PSub). The performance metric for the corresponding homogeneous structures is shown for comparison.} \label{Fig5}
\end{figure*}

In Ref.~\cite{Hussein_2015}, a PnC was used to form the PSub structure. The frequency band structure of a unit cell for this PnC is shown in~Fig.~\ref{Fig4}a. The finite extent of the PnC represents a symmetry breaking, or truncation, of an otherwise idealized PnC with an infinite extent. The symmetry breaking has been taken to our advantage as it created a \textit{truncation resonance} inside a band gap~\cite{Camley_1983,Davis_2011,Albabaa_2017,Bastawrous_2022,Albabaa_2022,Rosa_2022}, the first Bragg band gap for the unit cell considered. Associated with the truncation resonance, there is a phase change from positive to negative as the frequency is increased, allowing us to yield a negative value of $P$ with a high absolute value at frequencies higher than the truncation resonance frequency.
Furthermore, the negative $P$ properties extend over a relatively wide frequency range compared to what is produced by a standard structural resonance associated with, for example, a statically-equivalent homogeneous structure. The higher the value of $\left|P\right|$, the stronger the control, in the negative for stabilization or in the positive for destabilization, and the broader its frequency range, the more robust the control effect. The performance metric for a PnC-based PSub versus a PSub comprising a statically-equivalent homogeneous structure is shown and contrasted in Fig.~\ref{Fig5}a.

\subsection{Pass-band lowered-resonance approach}
\label{sec:PSubsPass}
As mentioned above, a PnC-based PSub must be relatively long to accommodate low-frequency instabilities. To mitigate this limitation, we demonstrate in this paper the concept of a subwavelength PSub using a locally-resonant elastic MM.~An elastic MM may be designed to feature a band gap in the subwavelength regime (i.e., where the wavelength of the elastic wave is larger than the unit-cell size of the periodic medium~\cite{Liu_2000}), as shown in Fig.~\ref{Fig4}b.~In this case, it is also possible to produce a finite structure with a truncation resonance inside a subwavelength band gap~\cite{xiao2013flexural,Sangiuliano_2020,Xia_2020,Park_2022}.~However, here we provide an alternative approach whereby the PSub resonance that we utilize is a \textit{sub-hybridization resonance}.~This global-structure resonance appears at a frequency lower than the subwavelength band gap, which otherwise would appear at a much higher frequency if the band gap did not exist. The performance metric for an MM-based PSub versus the same PSub without the local resonators is shown and contrasted in Fig.~\ref{Fig5}b.~The lowest resonance for the MM-based PSub is near 1550 Hz, whereas the lowest resonance for the same rod without the resonators is near 8000 Hz.

\section{Models and Methods}
\label{sec:1}
As described in Section~\ref{sec:PSubs}, a PSub is designed without the need for any coupled fluid-structure simulations$-$a trait that is indicative of the mechanistic nature of the theory of phononic subsurfaces. The theory entails producing four key plots that allow for full characterization of the properties of a given PSub configuration~\cite{Hussein_2015}.~The first is the dispersion curves (elastic band structure) for the unit cell from which the PSub is formed.~The second and third plots are the frequency-dependent amplitude and phase response of the PSub, with both the excitation and response being at the edge that will be exposed to the flow.~The fourth characterization calculation produces the frequency-dependent performance metric $P$ for the PSub, defined as the product of the amplitude and phase as mentioned earlier. These plots allow for prediction of the changes that will occur in the instability field in the flow when the PSub is installed. A simulation of the flow coupled to the PSub is then run only to verify and assess the performance.~In the simulation, the Navier-Stokes equations are solved simultaneously with Newton's second law governing the elastodynamic motion in the PSub, with appropriate boundary conditions applied at the fluid-structure interface.~This section briefly describes the models, numerical procedures, and physical parameters used throughout the paper.~The reader is referred to Ref.~\cite{Hussein_2015} for more details on the modeling and solution methods. 

\subsection{PSub model and analysis approach}
\label{PSubDispersionG}
All the PSub structures we investigate are modeled as one-dimensional (1D) linear elastic solid rods with a constant cross-sectional area, where the elastodynamic motion is governed by
\begin{equation}
    \rho_{\rm s} \ddot{\eta}=(E\eta_{,s}+C\dot{\eta}_{,s})_{,s}+f,
    \label{eq:structure}
\end{equation}
where the structure's axial spatial coordinate and time are denoted by $s$ and $t$, respectively, and $\rho_{\rm s}=\rho_{\rm s}(s)$, $E=E(s)$, $C=C(s)$, $\eta=\eta(s,t)$, and $f=f(s,t)$ represent the material density, elastic modulus, damping constant, longitudinal displacement, and external force, respectively. Differentiation with respect to position is indicated by $(.)_{,s}$, and the superposed single dot $\dot{(.)}$ and double dot $\ddot{(.)}$ denote the first and second time derivatives, respectively. \\ 
%\subsubsection{PSub unit-cell dispersion model} 
\indent The dispersion curves for a given PSub unit-cell configuration are obtained by setting the force $f$ to zero in Eq.~(\ref{eq:structure}) and applying Bloch's theorem ~\cite{Bloch_1929,Hussein_2009}; this yields a relationship between the frequency $\omega$ and the wavenumber $\kappa$ for longitudinal wave propagation along the axis of the rod. 
%\subsubsection{Finite PSub steady-state structural dynamics model} 
%\label{PSubFRF}
The amplitude and phase response of a finite version of the PSub composed of $n_c$ repeated unit cells are obtained by solving Eq.~(\ref{eq:structure}) as a boundary value problem. Free and fixed boundary conditions are chosen for the PSub top and bottom edges, respectively, i.e., $\eta_{,s}(0,t)=0$ and $\eta(l,t)=0$, where $l=n_c a_{\rm{UC}}$, and $a_\mathrm{UC}$ is the PSub unit-cell length. The top end is excited harmonically, i.e.,  $f(0,t) = \tilde{f}(0)\rm{e}^{\rm{i}\omega_{\mathrm{e}} \it{t}}$ and $f(s,t)=0$ for $s>0$, where $\omega_{\mathrm{e}}$ is the excitation frequency and $\tilde{f}$ is the amplitude of the forcing. The displacement response is given by $\eta(s,t) = \tilde{\eta}(s)\rm{e}^{\rm{i}\omega_{\mathrm{e}} \it{t}}$, where $\tilde{\eta}(s)$ is the amplitude of the response. The phase ${\phi}={\phi}(s)$ is formulated to span the range $-\pi/2\leq{\phi}\leq\pi/2$. 
%\subsubsection{Finite PSub wave propagation model} 
When running the coupled fluid-structure simulations, the PSub model must be treated as an initial boundary value problem where no assumptions are made for the displacement field's temporal dependency. As in the steady-state analysis, we set $f(s,t^*)=0$ for $s>0$ and the value of $f(0,t^*)$ is fed in from an integration of the pressure field exerted by the flow at each time step in the simulation covering the time interval $0 \leq t^* \leq t^*_{\rm T}$, where $t^*$ and $t^*_{\rm T}$ are the coupled simulation dimensional time and end time (in seconds), respectively.\\
%\subsubsection{Solid-domain numerical schemes}

 The PSub rod material/structure is numerically analyzed  using the finite-element (FE) method utilizing 1D 2-node iso-parametric elements \cite{HusseinJSV06}. Damping is introduced in the form of viscous proportional damping, which yields a unit-cell damping matrix defined as ${\mathbf{C}} = {q_1}{\mathbf{M}}  + {q_2}{\mathbf{K}}$, where $q_1$ and $q_2$ are damping constants and ${\mathbf{M}}$ and ${\mathbf{K}}$ denote the FE mass and stiffness matrices, respectively. The dispersion curves are obtained for values of wavenumber in the range $0 \leq \kappa \leq \pi/a_{\rm{UC}}$~\cite{Hussein_PRB_2009}. 
 The number of nodes in the unit cell is denoted by $n_\mathrm{m}$. For the finite version of the PSub, the number of nodes along the full structure is $n_s=n_c(n_{m}-1)+1$.~For the wave propagation simulation problem, the second-order Newmark time integration scheme is used with the dimensional time step increment $\Delta t^*$.~We use an implicit version of the scheme by selecting the parameters $\gamma=1/2$ and $\beta=1/4$ in the formulation provided in Ref.~\cite{Hussein_2015}. \\

\subsection{Model of unstable channel flow with PSub installed and simulation approach}
\label{FlowModel}
We examine spatially evolving instabilities in fully-developed incompressible plane channel flows, also  known as Poiseuille flows. The flow is driven by a mean pressure gradient between two parallel walls that are nominally rigid except for the region where the PSub is located. An exact solution of the Navier-Stokes equations gives the mean velocity for the flow field~\cite{navier1823memoire,stokes1845g}, which is considered the base inflow. The dynamic stability in this flow is governed by the Orr-Sommerfeld equation~\cite{Orr_I_1907,Orr_II_1907,Sommerfeld_1908}, which is obtained by linearizing the Navier-Stokes equations using the normal assumption.~As mentioned in Section~\ref{Sec:Introduction}, we consider TS waves as examples of two-dimensional (2D) evolving instabilities in parallel shear flows.~These waves are represented by growing eigensolutions of the Orr-Sommerfeld equation and have been observed in laboratory experiments for channel flows~\cite{Nishioka_JFM_1975} and earlier in boundary layer flows~\cite{Schubauer_1948,Klebanoff_JFM_1962}. In our coupled fluid-structure simulations, we superimpose a particular Orr-Sommerfeld unstable spatial mode at the channel inflow boundary. This causes an excitation of the parabolic base velocity, which provides a representative model of an unstable spatially-evolving transitional flow in a typical laboratory experiment.\\ 
%\subsubsection{Coupled fluid-structure simulation model}
\indent The simulations are based on the time-dependent, three-dimensional Navier-Stokes equations where the channel half-height $\delta$ and the centerline velocity 
$U_\mathrm{c}$ are used for nondimensionalization. The continuity and momentum equations are as follows, respectively
\begin{equation}
\label{eq:cont}
\frac{\partial u_i}{\partial x_i} = 0,
\end{equation}
\begin{equation}
\label{eq:NS}
\frac{\partial u_i}{\partial t} + \frac{\partial u_j u_i}{\partial x_j} = \frac{2}{Re}-\frac{\partial p}{\partial x_i} + \frac{1}{Re}\frac{\partial u_i}{\partial x_j x_j},
\end{equation}
where ${\bf u}(x,y,z,t)=(u,v,w)$ is the velocity vector with components in the streamwise $x$, wall-normal $y$, and the spanwise $z$, directions, respectively, and $p$ is the nondimensional pressure. Moreover, $Re=U_\mathrm{c}\delta/\nu_\mathrm{f}$ is the Reynolds number based on the centerline velocity, $\nu_\mathrm{f}$ is the kinematic viscosity, and $t$ (in this context) is the nondimensional time.~The ranges of the wall-normal and spanwise domains are $0\le y\le 2$ and $0\le z\le 2\pi$, respectively.~We decompose the velocity vector in Eqs.~\eqref{eq:cont} and \eqref{eq:NS} into ${\mathbf{u}} = {\bar{\mathbf{u}}} + {\mathbf {\hat{u}}}$, where ${\bf \bar{u}}$ is the mean flow component obtained by averaging ${\mathbf{u}}$ over a time range and ${\bf \hat{u}}$ is the perturbation (instability) component.~With this decomposition, $p = \bar{p} + \hat{p}$, where $\hat{()}$ represents the perturbation part of the flow. 

The initial and boundary conditions for the decomposed velocity field for an all-rigid-wall channel are
\begin{subequations}
	\begin{equation}
	\label{eq:bcgen}
	{\bf{u}}(x=0,y,z,t) = 
	{\bf{u_b}}(x=0,y,z,t)+{\bf{\hat{u}}}(x=0,y,z,t), 
	\end{equation}
	\begin{equation}
	\label{eq:bc1}
	{\bf u_b}(x=0,y,z,t) = \left(1-(1-y)^2, 0, 0\right), 
	\end{equation}
	\begin{equation}
	\label{eq:bc2}
	{\bf \hat{u}}(x=0,y,z,t) = A_\mathrm{2D}\text{Real}[{\bf u}_{\rm e2D}(y)e^{-{\rm i}\omega_{\mathrm{TS}}t}],
	\end{equation}
	\begin{equation}
	\label{eq:bc3}
	{\bf u}(x,y=0,2,z,t) = (0,0,0),
	\end{equation}
\end{subequations}
where $A_{\rm 2D}$ is the amplitude of the 2D perturbation, ${\bf u}_{\rm e2D}$ is the Orr-Sommerfeld eigenfunction we prescribe, and $\omega_\mathrm{{TS}}$ is the perturbation dimensionless frequency (which is a real quantity).~Only  the $\hat{u}$ and $\hat{v}$ components of ${\bf u}_{\rm e2D}$ are nonzero.~Furthermore, periodic boundary conditions are applied in the $z$ direction and a non-reflective buffer domain is added to the physical domain for the outflow boundary conditions~\cite{Hussein_2015,Dana91,Saiki93,Kucala14}.~The complex wave speed of the perturbation is defined as $c=-\omega_{\mathrm{TS}}/\alpha$ where $\alpha=\alpha_{\rm R}+{\rm i}\alpha_{\rm I}$ denotes complex wavenumber~\cite{Reynolds69}.~The perturbation grows in space when $-\alpha_{\rm I}>0$.

The PSub installation region covers a streamwise distance from $x_{\rm s}$ to $x_{\rm e}$ and extends uniformly across the entire spanwise direction. For the coupled simulations throughout this paper, $(\cdot)^*$ represents dimensional quantities, whereas the omission of the asterisk symbol denotes dimensionless flow quantities.~We define the dimensional wall pressure as $p^*_{\rm w}=\bar{p}\rho_{\rm f} U_{\rm c}^2$, where $\rho_{\rm f}$ is the fluid density and $\bar{p}$ is the averaged pressure between $x_{\rm s}$ and $x_{\rm e}$, respectively.~At every time step, this quantity is computed on the fluid-structure interface.~It acts on the top edge of the PSub as a force.~On the other hand, the resultant displacement $\eta(0,t^*)$ and velocity $\dot{\eta}(0,t^*)$ obtained from the time integration of the structure model is imposed as boundary conditions to the flow field at the interface such that~\cite{Hussein_2015}

\begin{subequations}
\label{eq:structbc}
	\begin{equation}
	\label{eq:structbcu}
	\hat{u}(x_{\rm s}\le x\le x_{\rm e},y=0,z,t)=-\frac{\eta(0,t^*)}{\delta}\frac{du_{\rm b}}{dy}, 
	\end{equation}
	\begin{equation}
	\label{eq:structbcv}
	\hat{v}(x_{\rm s}\le x\le x_{\rm e},y=0,z,t)=\frac{\dot{\eta}(0,t^*)}{U_\mathrm{c}}.
	\end{equation}
\end{subequations}
These boundary conditions ensure that the stresses and velocities match at
the fluid-structure interface and are valid when $\eta<<\delta$ is maintained throughout the computations.~Referred to as transpiration boundary conditions~\cite{Lighthill_1958,Sankar_1981}, Eqs.~(\ref{eq:structbcu}) and~(\ref{eq:structbcv}) are obtained by keeping the interface location fixed and retaining only the linear terms following a Taylor series expansion of the exact interface compatibility conditions.~Other boundary conditions have been examined by Barnes et al.~\cite{Barnes_2021} giving qualitatively similar results.~Given our assumption of small displacements, these fluid-structure interface boundary conditions allow wall motion predominantly along the wall-normal \it y\rm-direction, since $\dot{\eta}>>\eta$.~The spanwise velocity $w$ is zero at the interface.

For the flow field, the Navier-Stokes equations are integrated using a time-splitting scheme~\cite{Dana91,Saiki93,Kucala14} on a staggered structured grid system, in which the velocity components are computed at the edges, and the pressure is determined at the centers.~The wall-normal diffusion term is discretized by implementing the implicit Crank-Nicolson method, and the Adam-Bashforth scheme is used for an explicit treatment of all the other terms.~This numerical procedure was verified with the linear theory giving a maximum deviation of $0.05\%$ in the predicted perturbation energy growth~\cite{Kucala14}.~Since the equations for the fluid and the structure are inverted separately in the coupled simulations, a conventional
serial staggered scheme~\cite{Farhat_2000} is implemented to couple the two sets of time integration.
 
\subsection{Model parameters}
\label{sec:MP}
Table \ref{Table:phononic} lists the geometric parameters and material properties of the PSubs we examine in this paper.~For the PnC-based PSub, we select the values of 2.4 $\mathrm{GPa}$ and 1040 $\mathrm{kg/m^3}$ for the elastic modulus and density of ABS polymer, respectively, and the corresponding values of 68.8 $\mathrm{GPa}$ and 2700 $\mathrm{kg/m^3}$ for the Al.~The unit cell of the PnC rod  consists of two layers, aluminum (Al) and ABS polymer.~The PSub comprises 5 unit cells, each with a length of $a_{\rm PnC}=40$ cm (i.e., $l_\mathrm{PnC}=2$ m). In the FE analysis, each unit cell is discretized into 50 linear elements; hence, the structure has 250 degrees of freedom considering the fixed end at the bottom.

\begin{table}
\begin{center}
\caption{Geometric parameters and material properties of PSubs}
\label{tab:1}       % Give a unique label
%
% Follow this input for your own table layout
%
\begin{tabular}{p{2cm}p{3cm}p{3cm}p{2cm}p{3.4cm}}

\hline
Material & Volume Fraction & Elastic Modulus & Density & Damping Constant\\
  & ($\%$) & ($\mathrm{GPa}$) & ($\mathrm{kg/m^3}$) & ($q_1;q_2$)  \\
Aluminum & 90 & 68.8$-$70  & 2700$-$2710 & (0;6$e^{-9}$)\\
ABS & 10 & 2.4$-$3  & 1040$-$1200 & (0;6$e^{-8}$)\\
\hline
\end{tabular}
\label{Table:phononic}
\end{center}
\end{table}

The unit cell of the MM-based PSub consists of a homogeneous rod made out of ABS polymer and a local mass-spring resonator attached at the center.~This configuration may be realized in practice by, for example, a rod/beam structure with pillars periodically attached to represent the resonators~\cite{wu2008evidence,pennec2008low,Bilal_2013,Xiao_2013}.~We choose the elastic modulus and density of ABS polymer to be 3 $\mathrm{GPa}$ and 1200 $\mathrm{kg/m^3}$, respectively.~The unit cell has a length of $a_{\rm MM}=1$~cm, and the PSub is formed from either 5 ($l_{\rm MM}=5$~cm), 10 ($l_{\rm MM}=10$~cm), 15 ($l_{\rm MM}=15$~cm), or 20 ($l_{\rm MM}=20$~cm) unit cells. The resonator's mass and spring stiffness are tunable according to the target instability frequency. In the nominal case, the resonator's frequency is set to $f_\mathrm{res}=2000$ Hz. The resonator's point mass is set to be ten times higher than the total mass of the rod portion in the unit cell, $m_\mathrm{res}=10\times \rho_\mathrm{ABS}a_{\rm MM}$; this gives a resonator's stiffness equal to $k_\mathrm{res}=m_\mathrm{res}(2\pi f_\mathrm{res})^2$. The metamaterial unit cell is discretized into seven FE elements (including six rods and one mass-resonator elements); thus each unit cell has eight degrees of freedom including that of the resonator. A 5 unit-cell MM-based PSub would therefore have 35 degrees of freedom by applying fixed boundary conditions at the bottom.~The reader is referred to Ref. \cite{Khajehtourian_2014} for details on the dispersion behavior of this particular elastic metamaterial configuration.

The coupled fluid-structure simulations are based on ${Re}=7500$, incorporating an instability with nondimensional frequency $\omega_\mathrm{TS}=0.25$ and wavenumber $\alpha=1.0004-\mathrm{i}0.0062$, which corresponds to the least-attenuated eigenmode of the Orr-Sommerfeld equation.~Utilizing nondimensional analysis to simulate a given TS wave with a dimensional frequency $f_\mathrm{TS}={\omega_\mathrm{TS}}{U_{\rm c}}/{2\pi}{\delta}$ Hz, we vary the centerline velocity (velocity scale) ${U_{\rm c}}$ and half height of the channel (length scale) $\delta$ accordingly in the DNS code.~All simulations are done for liquid water for which the kinematic viscosity is $\nu_\mathrm{f}=1\times10^{-6}$ $\mathrm{m^2/s}$.~While not considered here, PSubs may also be designed for air by adjusting the elastic compliance of the PSub surface exposed to the flow.~The $\delta$ quantity varies between the different models examined. For example, a value of $\delta=4.23\times10^{-4}$ m is used for a PnC-based PSub targeting strong stabilization of instability at 1670 Hz (see details in Section~\ref{sec:ResPnC}) and $\delta=4.38\times10^{-4}$ m for an MM-based PSub comprising 5 unit cells and targeting strong stabilization of instability at 1550.3 Hz (see details in Section~\ref{sec:ResMM}). The corresponding centerline velocities for these PnC-based and MM-based PSub simulations are $U_{\rm c} = 17.72$ m/s and $U_{\rm c} = 17.11$ m/s, respectively. 

For all the MM-based and PnC-based PSub simulations, the dimension of the channel is fixed as $L_x=20\delta$, $L_y=2\delta$, and $L_z=2\pi \delta$. The fluid domain is discretized into $n_x=225$, $n_y=65$, and $n_z=8$ points in the streamwise, wall-normal, and spanwise directions, respectively. The length of the PSub interface (control surface) along the streamwise direction is approximately a quarter of the instability wavelength, $\lambda_\mathrm{TS}=2\pi\delta/\alpha_\mathrm{R}$. The front and end edges of the PSub interface in the streamwise direction are $x_s/\delta\approx6$ and $x_e/\delta\approx8$, respectively.~The dimensional time step $\Delta t^*$ is selected such that 2000 time steps cover a period of the instability wave. Specifically, $\Delta t^*=3\times10^{-7}$ s and $\Delta t^*=3.22\times10^{-7}$  s for the PnC-based and MM-based PSub simulations, respectively. The dimensional time integration step for the flow is the same as that for the PSub.~All the simulations are run for 3 million time steps until $t_{\rm end}^*\approx 1$ s where $t_{\rm end}^*$ is the dimensional time at the end of the coupled fluid-structure simulations. The averaging time window for adequately capturing the relevant statistics for the various cases is chosen to begin when the simulation has become quasi-steady, i.e., the effect of the initial conditions has faded, and to extend sufficiently long to cover approximately 1000 TS wave periods.~The buffer region is sized to 40$\%$ of the channel length ending at the outlet~\cite{Hussein_2015}. All the simulations were executed on the RMACC supercomputer Summit at the University of Colorado implementing parallel computation. 

\section{Results}
\label{sec:Res}

We now examine the detailed characteristics and actual performance from coupled fluid-structure simulations of the two types of PSubs considered in Figs.~\ref{Fig2},~\ref{Fig4} and~\ref{Fig5}. 

\subsection{PnC-based PSub}
\label{sec:ResPnC}
\begin{figure*} [t!]
	\includegraphics[width=1\columnwidth]{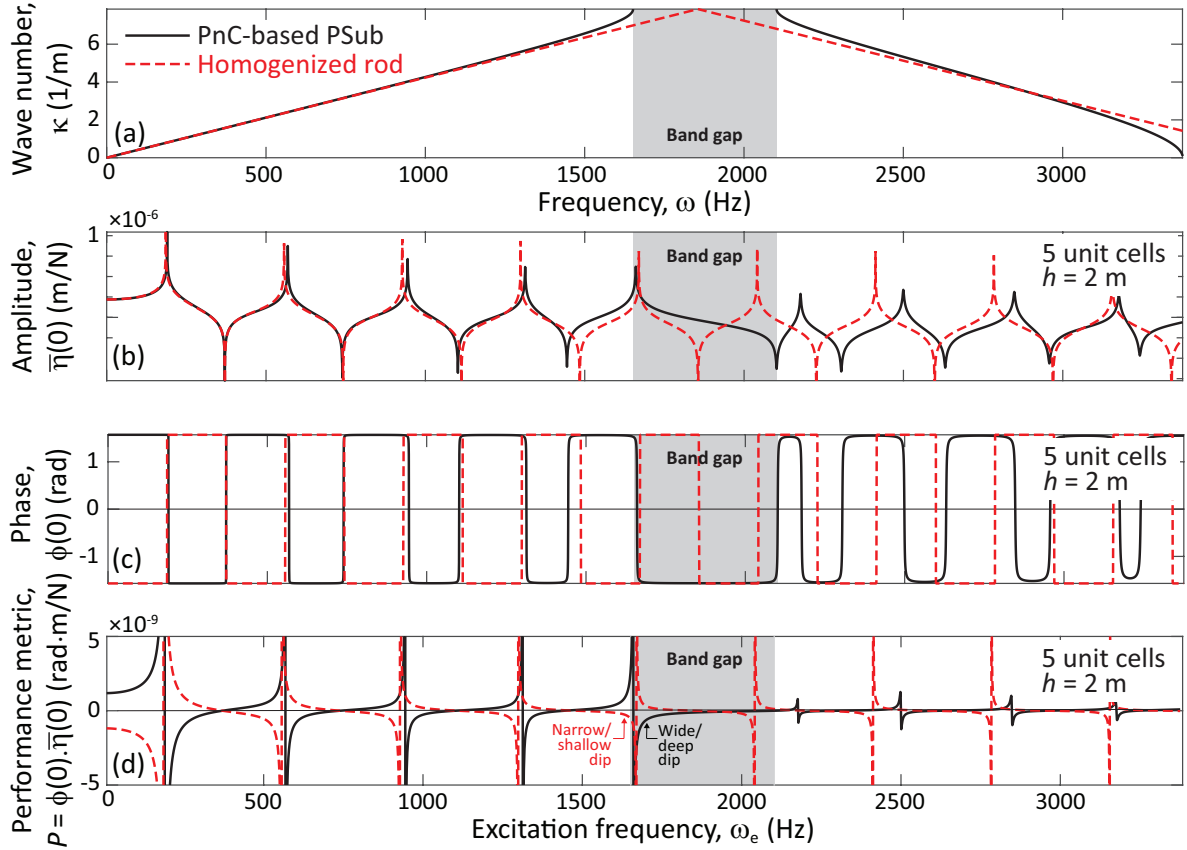}%
	\caption{Four key characterization plots that form the foundation of the PSubs theory: (a) Dispersion curves for a unit cell from which the PSub is formed. Steady-state vibration (b) amplitude and (c) phase response of the PSub top edge when harmonically excited at the same location.~(d) Performance metric obtained by multiplying the amplitude by the phase. The phase is between the force and the displacement at the PSub top edge. All plots are obtained by analyzing a stand-alone FE model of the PSub without yet coupling to the flow. These results are for the PnC-based PSub with a 40-cm long unit cell.} \label{Fig6}
\end{figure*}
The four key characterization plots for the PnC-based PSub configuration whose geometric and material properties are given in Section~\ref{sec:MP} are shown in Fig.~\ref{Fig6}. This structure is identical to that investigated in Ref.~\cite{Hussein_2015} which was designed for a TS instability with a frequency of 1690 Hz, except here it comprises five unit cells instead of 10.~The band structure pertaining to the PSub unit cell features a band gap, as shown by the grey region throughout the four plots. A truncation resonance appears inside the band gap at 1660.3 Hz for five unit cells; and, as shown in the third plot, the phase turns from positive (in-phase) to negative (out-of-phase) at that frequency and stays negative until the next resonance. This, in turn, gives a value of $P$ that is positive at pre-resonance and negative at post-resonance, as shown in the fourth plot. Both the amplitude and phase quantities are determined from isolated steady-state harmonic frequency response analysis of a 5-unit-cell long version of the PSub with fixed support at the bottom, as described in Section~\ref{PSubDispersionG}. This contrasts with Ref.~\cite{Hussein_2015} where the phase spectrum was obtained by running long-time simulations. For comparison, the characterization curves of the statically equivalent homogeneous structure are superimposed in all plots. It is noticeable that the distance between the resonances and the range of the negative phase for the homogeneous structure near the TS wave frequency peak is markedly narrower than that of the PnC-based PSub. Consequently, the dip in the $P$ curve near the TS wave frequency is both wider (broader) and deeper (higher in absolute value) for the PnC-based PSub compared to the corresponding homogenized structure, as marked in Fig.~\ref{Fig6}d.~This advantage is present for both the functions of stabilization and destabilization. 

\begin{figure*} [t!]
	\includegraphics[width=1\columnwidth]{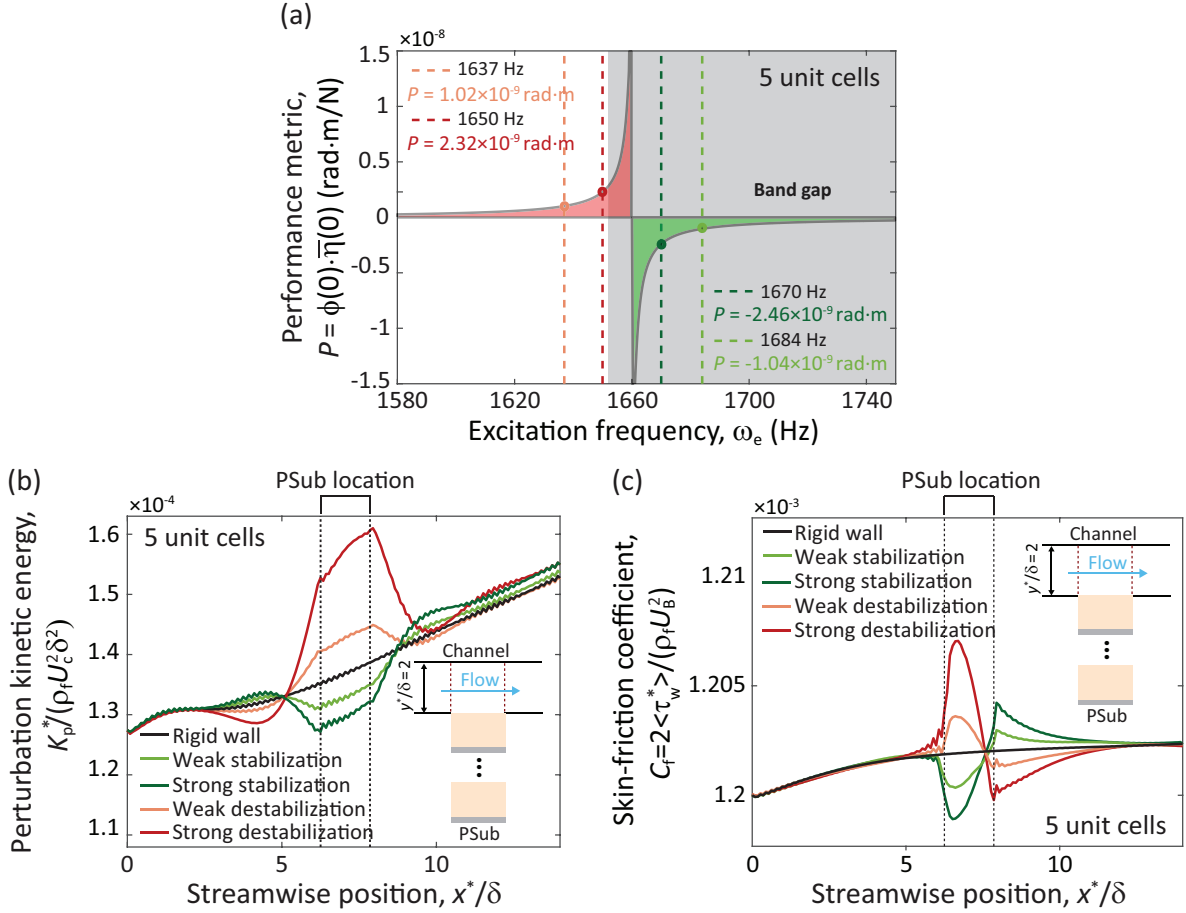}%
\caption{Demonstration of PnC-based PSub performance for both flow stabilization and destabilization.~(a) Performance metric curve (grey) and four vertical lines respectively representing four different instability waves investigated (each characterized by a frequency as indicated).~Green and red regions quantify the intensity and frequency breadth of the stabilization and destabilization capacity of the PSub; grey region represents frequency range of band gap.~Time-integrated (b) kinetic energy of the flow perturbation (instability) and (c) skin-friction coefficient as a function of streamwise position for each of the four cases as obtained from coupled flow-PSub simulations.~The PSub location spans the distance between the two dashed lines as indicated.~The responses quantitatively correlate with the frequency-performance metric intersection values in (a), indicating a perfect prediction of PSub performance. } \label{Fig7}
\end{figure*}

In Fig.~\ref{Fig7}a, we show a portion of the $P$-function again and mark the frequency values of four different TS wave instabilities. The first from the left (light orange line) is at 1637 Hz, which intersects the performance metric curve at a relatively low positive value ($P=1.02\times10^{-9}$ rad$\cdot$m/N)$-$indicating the ability to trigger \textit{weak destabilization} once the PSub is applied to a flow carrying an instability at this particular frequency. The second line from the left (dark red) is at 1650 Hz and can be seen to intersect the $P$ curve at a higher positive value ($P=2.32\times10^{-9}$ rad$\cdot$m/N), indicating the ability to cause~\textit{strong destabilization}.~The third frequency (dark green line) has a value of 1670 Hz; this intersects with the $P$ curve at a relatively high negative value ($P=-2.46\times10^{-9}$ rad$\cdot$m/N) which would bring rise to \textit{strong stabilization}. Lastly, the fourth vertical line (light green curve) corresponds to a TS wave with a frequency of 1684 Hz; this intersects with the performance metric curve at a lower negative value ($P=-1.04\times10^{-9}$ rad$\cdot$m/N) which would cause \textit{weak stabilization}.~Figure~\ref{Fig7}b shows the actual performance of the PSub in passively controlling each of these instabilities as seen from four separate coupled fluid-structure simulations.~To serve as a reference case, a fifth simulation is conducted with no PSub installed (i.e., the flow is exposed to a rigid wall all along) with a TS wave at 1660.3 Hz, corresponding to the center between the resonance and anti-resonance peaks in the PSub performance metric shown in Fig.~\ref{Fig7}a.~The figure shows a time-averaged quantity of the kinetic energy of the perturbation velocity field plotted as a function of the streamwise position. The perturbation kinetic energy $K_\mathrm{p}^{*}$, in unit of ${\rm J}/{\rm m}$, is defined as 

\begin{equation}
K_{\mathrm{p}}^*\left(x^*\right)=\rho_{\mathrm{f}} \int_0^{L_z} \int_0^{2 \delta} \frac{1}{2}\left(\left\langle\hat{u}^{* 2}\right\rangle+\left\langle\hat{v}^{* 2}\right\rangle+\left\langle\hat{w}^{* 2}\right\rangle\right) \mathrm{d} y^* \mathrm{~d} z^*,
    % K_\mathrm{p}^{*}(x^{*})=\rho_\mathrm{f} \int_{0}^{L_z} \int_{0}^{2\delta}\frac{1}{2}\left(\langle {\hat{u}^{*}}^{2} \rangle + \langle {\hat{v}^{*}}^{2} \rangle + \langle {\hat{w}^{*}}^{2} \rangle \right) \mathrm{d} y^{*} \mathrm{d} z^{*},
\end{equation}

\noindent where $\hat{u}^{*}$, $\hat{v}^{*}$, and $\hat{w}^{*}$ are the perturbation velocity components in the streamwise, wall-normal, and spanwise directions, respectively.~The symbol $\langle \cdot \rangle$ denotes time-averaged quantities.~The channel flow characteristics are expected to be nonhomogeneous along the streamwise direction due to the presence of the instability and PSub.~It is clearly observed from Fig.~\ref{Fig7}b that the $K_\mathrm{p}^{*}$ of the instability field rises above the reference rigid-wall case for the destabilization cases and falls under it for the stabilization cases, and this rise or fall takes place exactly where the PSub is installed (as indicated by the two vertical lines). Furthermore, the intensity of the rise or fall of $K_\mathrm{p}^{*}$ is consistent with the absolute value of the performance metric at the frequency intersects in Fig.~\ref{Fig7}a, where a small value of $|P|$ correlates with a weak change in $K_\mathrm{p}^{*}$ and a large value of $|P|$ correlates with a strong change in $K_\mathrm{p}^{*}$. We also observe that the $K_\mathrm{p}^{*}$ levels return to nearly the same level of the reference rigid-wall case downstream to the PSub, which is a desired outcome as it indicates precise local control of the instability field.~The stronger the stabilization or destabilization within the PSub region, the larger the offset of $K_\mathrm{p}^{*}$ in the far downstream region compared to the rigid-wall case. 

In Fig. \ref{Fig7}c, we present the skin-friction coefficient calculated at the bottom wall of the channel where the PSub is installed.~The skin-friction coefficient $C_{\rm f}$ for channel flows is defined as
\begin{equation}
    C_{\rm f}(x^{*}) = \frac{\langle\tau_{\rm w}^{*}\rangle}{\frac{1}{2}\rho_{\rm f}U_{\rm B}^2}
\end{equation}
where $\langle\tau_{\rm w}^{*}\rangle=\left[\mu_\mathrm{f}\frac{\partial \langle u^{*} \rangle}{\partial y^{*}}- \rho_\mathrm{f}\langle \hat{u}^* \hat{v}^*\rangle\right]_{y^{*}=0}$ is the wall mean shear stress, $\mu_\mathrm{f}$ is the fluid's dynamic viscosity, and $U_\mathrm{B}$ is the bulk velocity.~The mean shear stress at the wall was computed using a polynomial fit.~We observe that the skin-friction coefficient decreases in the stabilization cases and increases in the destabilization cases within the PSub control region.~The behavior of the skin friction is, therefore, compatible with what we observe for the perturbation kinetic energy in~Fig.~\ref{Fig7}b. In the region where the perturbation kinetic energy decreases, the wall mean shear stress $\langle\tau_{\rm w}^{*}\rangle$ also reduces, resulting in a drop in the skin-friction coefficient values and vice versa for the destabilization cases.~This reduction (or enhancement) for the stabilization (or destabilization) is mild (less than 0.5$\%$) because the PSub region is relatively small, and the TS wave examined is growing slowly and represent a small linear perturbation in the flow. Nevertheless, the PSub is shown to influence the flow precisely as desired, and stronger influence on the skin friction will be achieved with greater area coverage by PSubs acting on more dominant instability fields.~A PSub designed to exhibit even higher values of $|P|$ at the instability frequency will also cause stronger changes to the skin-friction coefficient.
\begin{figure*} [t!]
	\includegraphics[width=1\columnwidth]{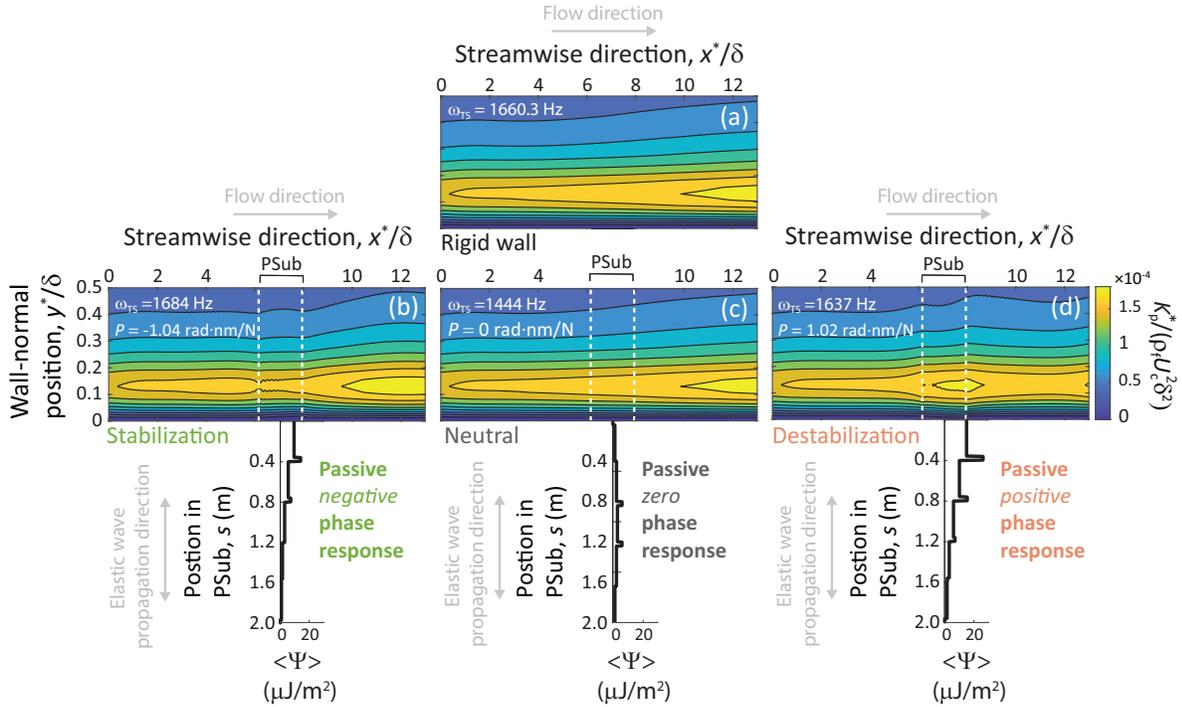}%
	\caption{Synchronized passive phased response and energy exchange between PnC-based PSub and flow perturbation (instability) field. The colored contours show the time-averaged spatial distribution of the perturbation kinetic energy within the flow.~The black curves represent the time-averaged total elastodynamic energy in the PSub. Stabilization ($P<0$) and destabilization ($P>0$) cases are shown in (b) and (d), respectively, whereas a neutral case ($P=0$) is shown in (c). The rigid-wall case is shown in (a) as a reference.}\label{Fig8}
\end{figure*}

The time-averaged spatial distribution of $K_\mathrm{p}^{*}$ over both the $x$ and $y$ directions is shown in Fig.~\ref{Fig8}, for the rigid-wall (Fig.~\ref{Fig8}a), weak stabilization (Fig.~\ref{Fig8}b), and weak destabilization (Fig.~\ref{Fig8}d) cases, respectively.~Figure (Fig.~\ref{Fig8}c) examines a case for $\omega_{\mathrm{TS}}=1444$ Hz, which corresponds to $P=0$ thus offering a neutral effect.~For Figs.~\ref{Fig8}b-d, we also show the corresponding time-averaged quantities of the total elastodynamic energy within the PSub, defined as $\Psi(s,t^*) = \frac{1}{2} \left( E \eta^2  +  \rho_{s} \dot{\eta}^2 \right)$, as obtained simultaneously from the same coupled simulations.~The peaks in the PSub total energy plots correspond to the regions occupied by the aluminum layers, where the speed of sound is higher than that of the ABS polymer layers.~We observe the total energy profile for the stabilization case (Fig.~\ref{Fig8}b) to be lower overall than that of the destabilization case (Fig.~\ref{Fig8}d), which is expected because the former admitting at out-of-phase (cancelling) wave motion across the fluid-structure assembly, and the latter is admitting in-phase (adding up) wave motion.~The total elastodynamic energy in the neutral case (Fig.~\ref{Fig8}c) is very small (almost negligible) because the PSub response amplitude at that frequency is zero (hence $P = 0$), thus preventing the system from experiencing any substantial fluid-structure interaction.~The results of Fig.~\ref{Fig8} demonstrate, most explicitly, a remarkable passive synchronization of response across both the PSub structure and the coupled flowing fluid. \\
\begin{figure*} [b!]
	\includegraphics[width=1\columnwidth]{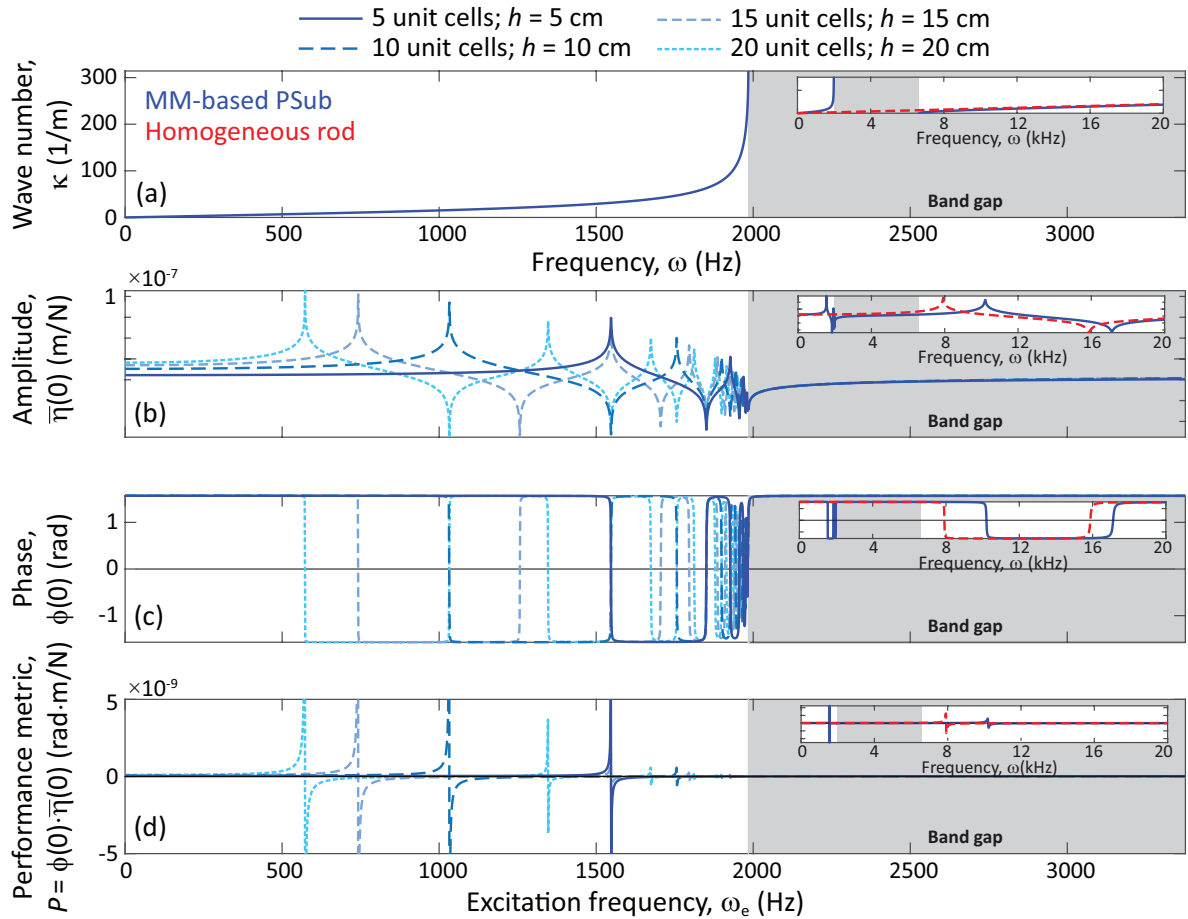}%
	\caption{Four key characterization plots for MM-based PSubs: (a) Dispersion curves for a unit cell from which each MM-based PSub is formed. Steady-state vibration (b) amplitude and (c) phase response at top edge of a 5-, 10-, 15-, or 20-unit-cell long PSub, all when harmonically excited at the same location.~(d) Performance metric obtained by multiplying the amplitude by the phase. The phase is between the force and the displacement at the PSub top edge.~Insets show the plots over an extended frequency range for the 5 unit-cell case and its corresponding homogeneous rod.~All plots are obtained by analyzing a stand-alone FE model of the PSub without yet coupling to the flow.~The unit cell of the MM-based PSub is 1-cm long.} \label{Fig9}
\end{figure*}
\begin{figure*} [b!]
	\includegraphics[width=1\columnwidth]{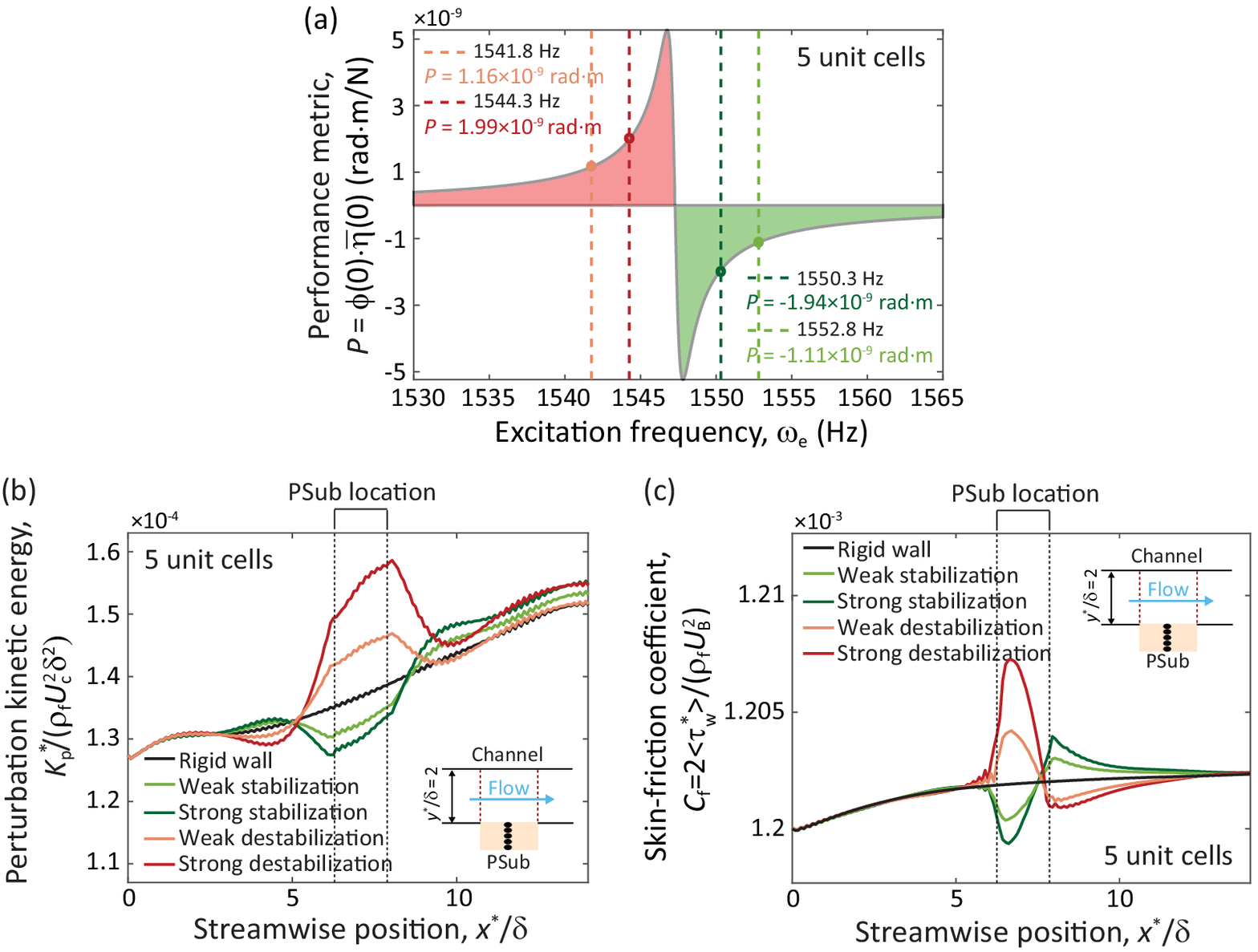}%
	\caption{Demonstration of MM-based PSub performance for flow stabilization and destabilization.~(a) Performance metric curve (grey) and four vertical lines respectively representing four different instability waves investigated (each characterized by a frequency as indicated).~Green and red regions quantify the intensity and frequency breadth of the stabilization and destabilzation capacity of the PSub.~Time-averaged (b) kinetic energy of the flow perturbation (instability) and (c) skin-friction coefficient as a function of streamwise position for each of the four cases as obtained from coupled flow-PSub simulations.~The PSub location spans the distance between the two dashed lines as indicated.~The responses quantitatively correlate with the frequency-performance metric intersection values in (a), indicating perfect prediction of PSub performance.} \label{Fig10}
\end{figure*}

\subsection{MM-based PSub}
\label{sec:ResMM}
As described in Section~\ref{sec:PSubsPass}, the MM-based PSub design approach utilizes a pass-band resonance that has been lowered in its frequency value due to the presence of a locally resonant hybridization band gap. The unit-cell dispersion diagram of the MM-based PSub configuration whose material and geometric properties are given in Section~\ref{sec:MP} is shown in Fig.~\ref{Fig4}b.~The dispersion curves for the same homogeneous rod but without the resonators are also shown for comparison.~Given that this PSub configuration comprises a slender homogeneous rod with a periodic arrangement of spring-mass resonators, a local-resonance band gap may be tuned to a target resonator frequency by simply adjusting the spring constant and/or mass value. As mentioned in Section~\ref{sec:MP}, we select a target resonator frequency of 2000 Hz and a resonator-to-rod mass ratio of 10; this generates a band gap centered at 4302.3 Hz.~Figure~\ref{Fig9}a shows the same dispersion diagram as the one shown in Fig.~\ref{Fig4}b, but in a rotated view and expressed in terms of dimensional frequency, and Figs.~\ref{Fig9}b,~\ref{Fig9}c, and~\ref{Fig9}d show the three remaining characterization plots, for each of a 5-, 10-, 15-, and 20-unit-cell long PSub.~We utilize the following subwavelength resonance frequency for each case: 1547.3 Hz (5-unit-cell PSub), 1032.4 Hz (10-unit-cell PSub), 742.2 Hz (15-unit-cell PSub), and 573 Hz (20-unit-cell PSub). The longer the PSub we can afford to install, the lower the frequency we can target for TS wave stabilization or destabilization for a given MM unit-cell configuration.~As seen in Fig.~\ref{Fig9}b, the PSub unit-cell band gap enables the generation of several structural resonances at frequencies lower than the band gap, which itself is already in the subwavelength regime. In particular, for the shortest PSub with 5 unit cells, we employ the resonance at 1547.3 Hz for our flow control objective. Similar to the results for the PnC-based PSub shown in Fig.~\ref{Fig7}, when comparing Fig.~\ref{Fig10}a with Fig.~\ref{Fig10}b we observe a direct correlation between the $P$ value at the intersection with the TS wave frequency and the corresponding actual $K_\mathrm{p}^{*}$ performance in the flow simulation.~Once again, we observe a perfect~\textit{a priori} prediction of whether the TS wave stabilizes or destabilizes, and at what level in each case.~Furthermore, similar to the PnC-based PSub cases, all the reductions in $K_\mathrm{p}^{*}$ take place exactly where the PSub is placed, and, favorably, the $K_\mathrm{p}^{*}$ levels return to nearly the same level of the reference rigid-wall case downstream to the PSub.~Figure.~\ref{Fig10}c displays the corresponding skin-friction coefficient calculated at the bottom wall of the channel, with qualitatively similar results to the PnC-based PSub results shown in Fig.~\ref{Fig7}c.~The rigid-wall case here is taken for a TS wave at 1547.3 Hz, corresponding to the center between the resonance and anti-resonance peaks in the PSub $P$ metric shown in Fig.~\ref{Fig10}a.

\begin{figure*} [t!]
	\includegraphics[width=1\columnwidth]{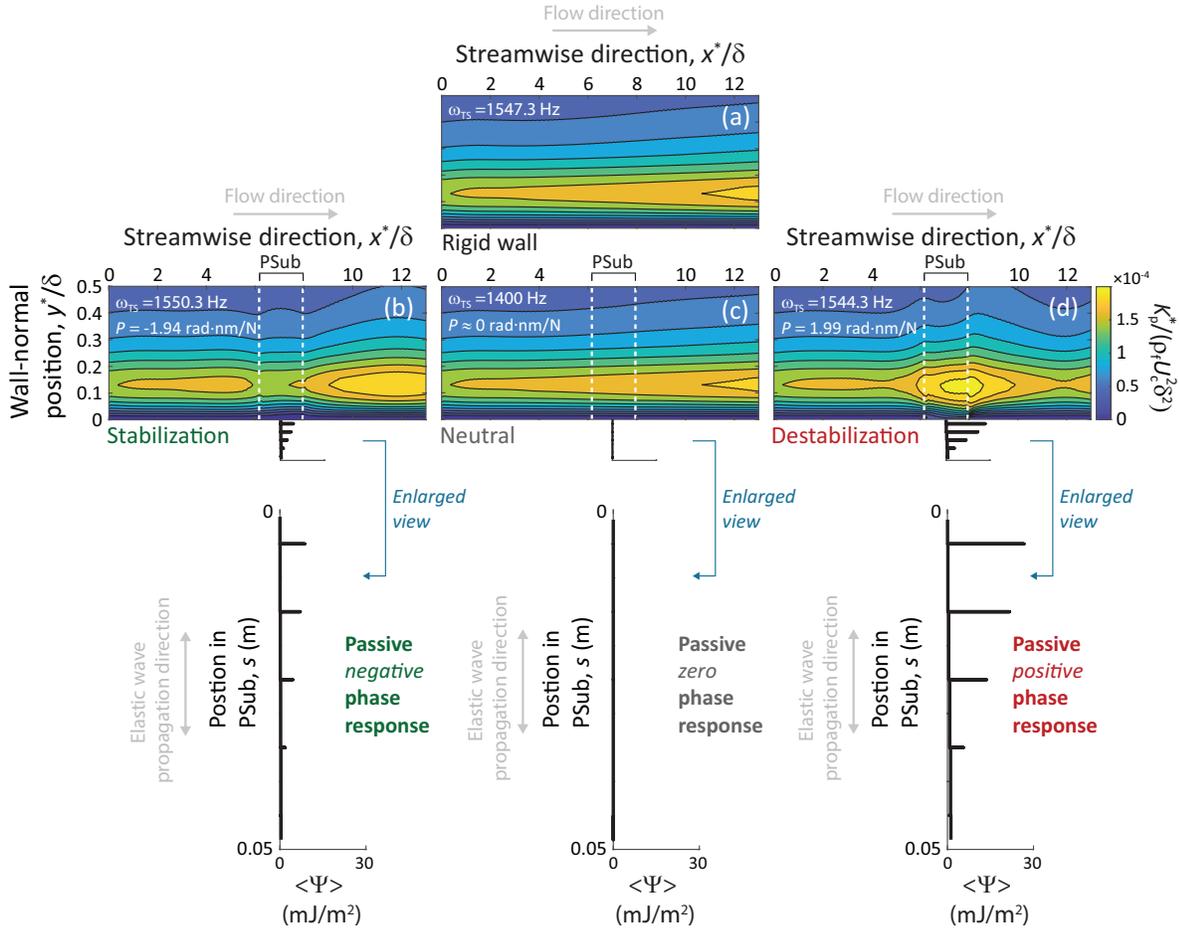}%
	\caption{Synchronized passive phased response and energy exchange between MM-based PSub and flow perturbation (instability) field. The colored contours show the time-averaged spatial distribution of the perturbation kinetic energy within the flow.~The black curves represent time-averaged total elastodynamic energy in the PSub, with horizontal lines indicating the total energy level of the resonating masses.~Stabilization ($P<0$) and destabilization ($P>0$) cases are shown in (b) and (d), respectively, whereas a neutral case ($P\approx0$) is shown in (c). The rigid-wall case is shown in (a) as a reference.}\label{Fig11}
\end{figure*}
\begin{figure*} [t!]
	\includegraphics[width=1\columnwidth]{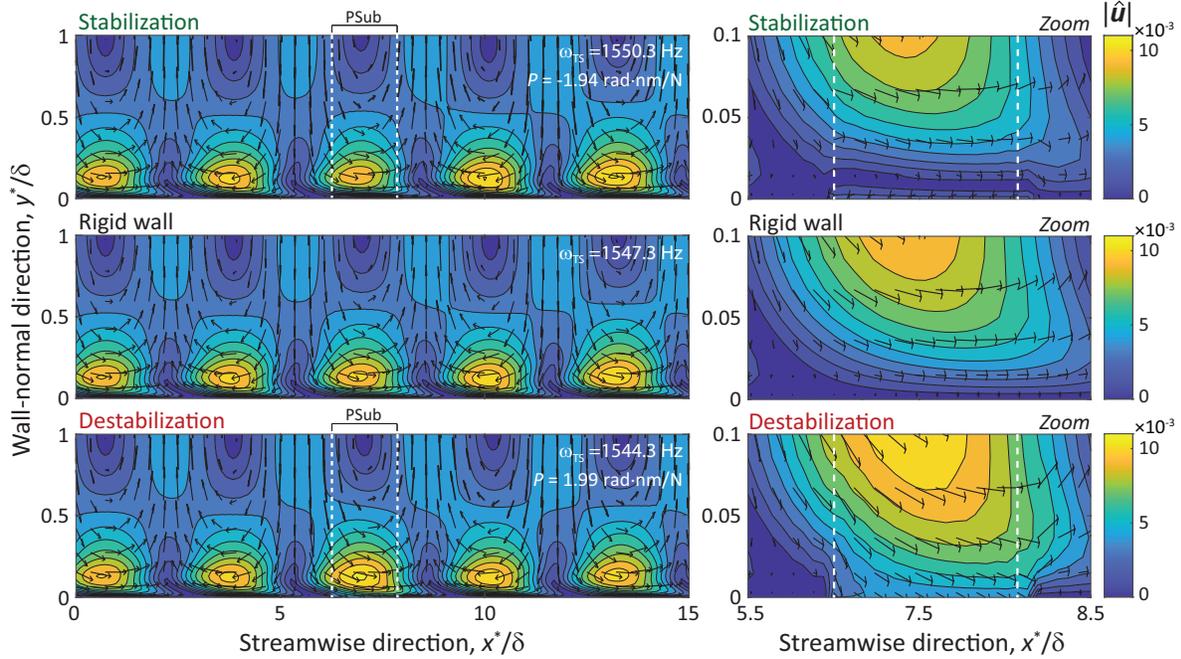}%
\caption{Instantaneous vector field of the perturbation (instability) velocity component $\hat{\mathbf{u}}$.~In the background, the resultant magnitude of the perturbation velocity field $|\hat{\mathbf{u}}|$ is also plotted.~Both are plotted along the $z=\pi$ plane.~In the top and bottom panels, the strongest stabilization and destabilization cases of Fig.~\ref{Fig10} are shown, respectively.~The PSub location spans the distance between the two white dashed lines as indicated.~The corresponding all-rigid-wall case is shown in the middle panel for comparison. Close-up views are shown in the right column.} \label{Fig12}
\end{figure*}

\begin{figure*} [t!]
	\includegraphics[width=1\columnwidth]{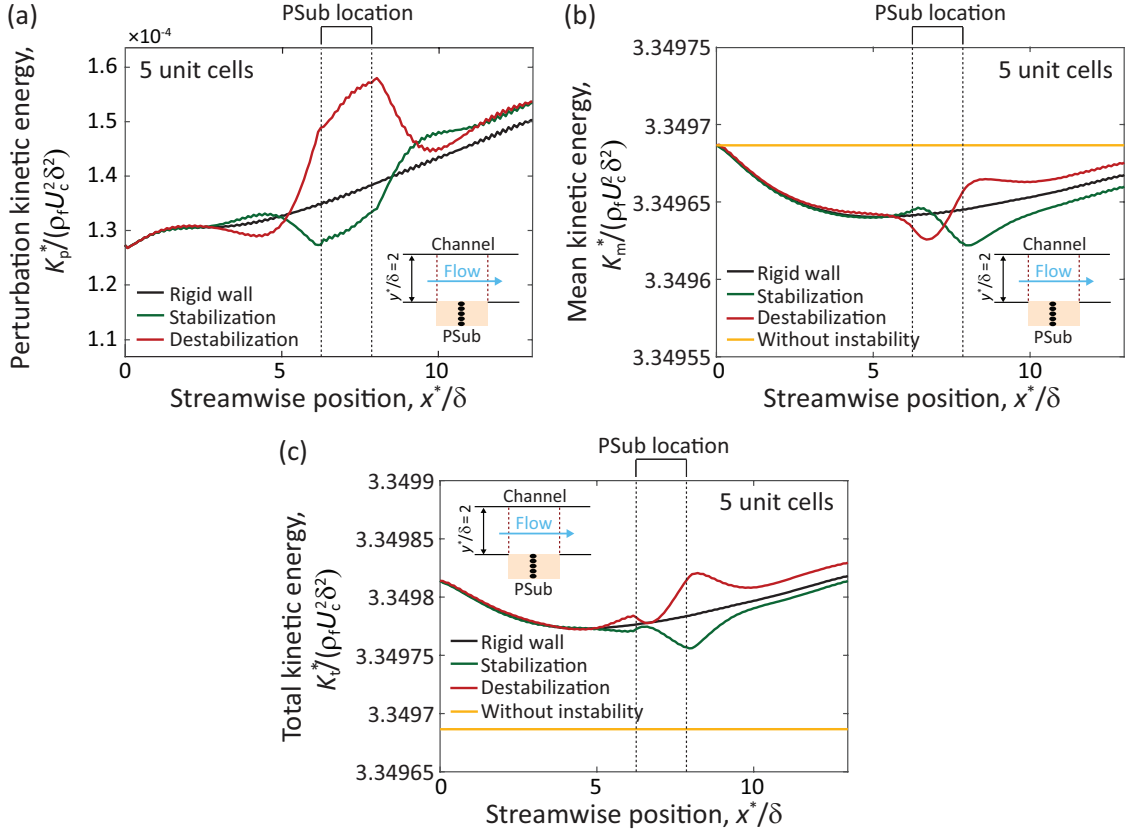}%
	\caption{Decomposition of flow kinetic energy.~Time-averaged kinetic energy of (a) perturbation (instability) component, (b) mean flow component, and (c) summation of both components, i.e., total kinetic energy. All results are for the MM-based PSub examined in Fig.~\ref{Fig10}; only the strongest stabilization and destabilization cases are shown.~The PSub location spans the distance between the two white dashed lines as indicated.~In each of (b) and (c), the corresponding kinetic energy curve for the same channel flow without the presence of an instability is shown (light orange). In all sub-figures, the case with the $\hat{u}$ (streamwise) component of the fluid-structure interface boundary conditions not applied (i.e., replacing Eq.~(\ref{eq:structbcu}) with $\hat{u}(x,0,z,t)=0$) is shown in dashed lines. } \label{Fig13}
\end{figure*}
\begin{figure*} [t!]
	\includegraphics[width=0.8\columnwidth]{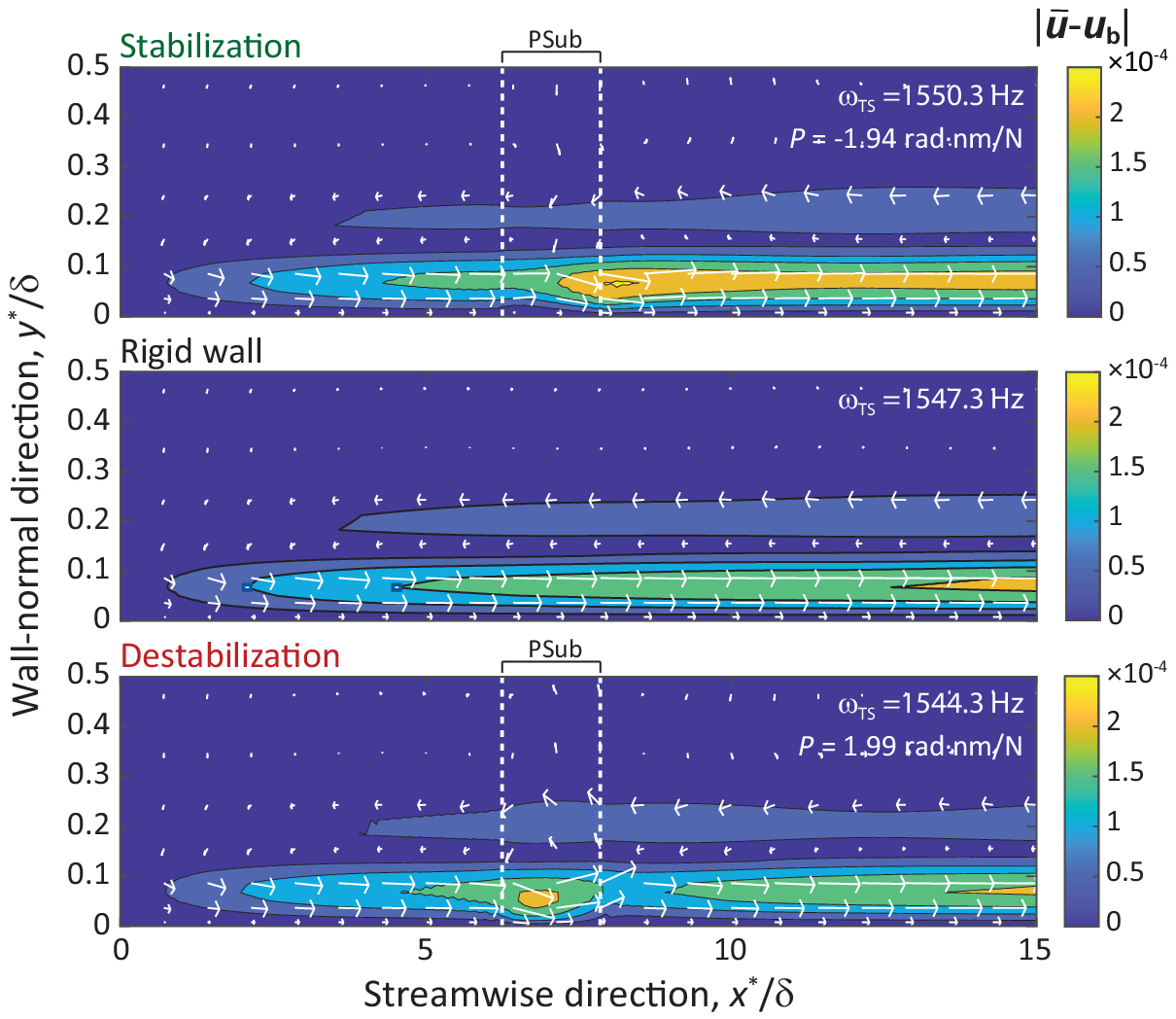}%
	\caption{Instantaneous vector field of the mean-flow velocity component with the base-flow velocity subtracted. i.e.,  $\bar{\mathbf{u}}-\mathbf{u_{\mathbf{b}}}$.~In the background, the resultant magnitude of this quantity, i.e., $|\bar{\mathbf{u}}-\mathbf{u_{\mathbf{b}}}|$, is also plotted.~Both are plotted along the $z=\pi$ plane.~In the top and bottom panels, the strongest stabilization and destabilization cases of Fig.~\ref{Fig10} are shown, respectively.~The PSub location spans the distance between the two white dashed red lines as indicated.~The corresponding all-rigid-wall case is shown in the middle panel for comparison.} \label{Fig14}
\end{figure*}
The time-averaged spatial distribution of $K_\mathrm{p}^{*}$ in the flow and the corresponding time-averaged total elastodynamic energy $\Psi(s,t^*)$ within the MM-based PSub are shown in Fig.~\ref{Fig11}. The rigid-wall (Fig.~\ref{Fig11}a), strong stabilization (Fig.~\ref{Fig11}b), and strong destabilization (Fig.~\ref{Fig11}d) cases, as well as a neutral case at 1400 Hz where $P\approx0$ (MM-based PSub does not generate zero $P$ before resonance frequency) (Fig.~\ref{Fig11}c), are shown. The black horizontal lines represent the total energy level of the locally resonating masses depicted in Fig.~\ref{Fig2}b.~In analogy to in Fig.\ref{Fig8}, we observe the energy in the resonators for the stabilization case (Fig.~\ref{Fig11}b) to be lower overall than that of the destabilization case (Fig.~\ref{Fig11}d), and also note that the the neutral case (Fig.~\ref{Fig11}c) experience very small (almost negligible) energy in the resonators.~As in Fig.~\ref{Fig8} for a PnC-based PSub, the results of Fig.~\ref{Fig11} for a MM-based PSub demonstrate a holistic synchrony in the coupled fluid-structure interaction response, and exactly consistent with the corresponding $P$ value in each case. 

Figure~\ref{Fig12} provides a contour plot of the absolute value of the instantaneous velocity perturbation for the strong stabilization and destabilization cases and the rigid-wall case for comparison.~A snapshot of the instantaneous vector field of the perturbation velocity is overlaid in each subfigure.~It is clear that at the PSub region, the stabilization case attains the lowest value of $|\mathbf{\hat{u}}|$ (smallest and least bright yellow spot), followed by the rigid-wall case (where these is no PSub), and then the destabilization case.~Consistent with this pattern, the perturbation velocity vector field experiences the smallest wall-normal components near the wall at the PSub region for the stabilization case, also followed by the rigid-wall case, and then the destabilization case.~Small wall-normal components compared to the rigid-wall case are indicative of coherent wave cancellation due to the presence of a stabilizing PSub.~In contrast, relatively large wall-normal components near the wall are indicative of destructive interference from a destabilizing PSub.

In Fig.~\ref{Fig13}, we examine the exchange of energy within the flow.~With no PSub installed, Fig.~\ref{Fig13}b shows that the mean-flow kinetic energy drops at the upstream region of the channel as the perturbation kinetic energy grows and acquires energy from the mean flow.~The trend eventually reverses when the mean flow begins to experiences structural changes itself as it carries a growing instability.~The time-averaged perturbation kinetic energy for the strong stabilization and destabilization cases are shown, again, in Fig.~\ref{Fig13}a and contrasted with the corresponding mean-flow component that is plotted in Fig.~\ref{Fig13}b.~The sum of both components is given in Fig.~\ref{Fig13}c.~The changes incurred in the controlled mean-flow component are very small due to the small magnitude of the perturbation, but nevertheless reveal valuable qualitative information.~In the presence of a PSub, we observe a short rise (fall) in the mean-flow kinetic energy near the upstream border of the PSub while the perturbation kinetic energy drops (rises) for stabilization (destabilization).~Subsequently, as the perturbation kinetic energy profile reverses direction, a corresponding opposite change in direction is seen in the mean-flow kinetic energy profile.~These trends confirm the energy exchange mechanisms depicted in the Fig.~\ref{Fig3} schematic. 

Fig.~\ref{Fig14} examines the influence on the mean flow from a contour diagram perspective.~In this figure, the base flow field is subtracted from the mean flow field yielding a $\mathbf{\bar{u}-u_{\rm b}}$ vector field which is plotted in 2D space.~Furthermore, the corresponding time-averaged quantity $|\mathbf{\bar{u}-u_{\rm b}}|$ is mapped out using color contours.~First we observe in the rigid-wall case that the velocity vectors points backwards (opposite to flow direction) near the middle of the half-channel, and, conversely, point forwards near the wall.~This pattern reveals that the instability is causing the mean-flow velocity profile to shorten and broaden, demonstrating very early traits of birth of transition to turbulence.~In the stabilization and destabilization plots, we observe an increase (decrease) in the mean-flow resultant amplitude and a pointing up (down) of the arrows near the wall for the cases of stabilization (destabilization). This reveals slower (faster) transition process in comparison with the rigid-wall case.~This adds further evidence of the phased energy exchange mechanisms described and discussed earlier.

\begin{figure*} [t!]
	\includegraphics[width=1\columnwidth]{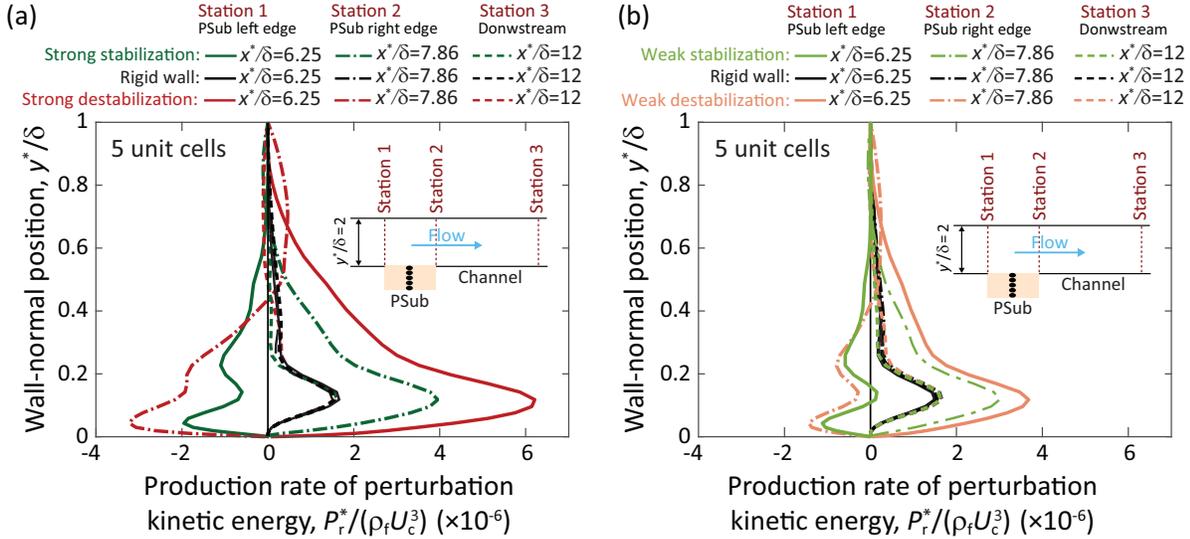}%
	\caption{Production of flow perturbation (instability) kinetic energy in channel with MM-based PSub for (a) strong and (b) weak stabilization or destabilization as a function of the wall-normal direction, $y^*/\delta$ at three streamwise positions (denoted measuring stations). Station 1 is located at the left edge position (beginning) of the PSub, Station 2 at the right edge position (end) of the PSub, and Station 3 at a far position downstream from the PSub.~A schematic of the PSub-installed channel with the station locations marked is shown in the insets.} \label{Fig15}
\end{figure*}

To further examine the underlying anti-resonance and resonance mechanisms within the flow, we compute the production rate of the perturbation  energy $P_{\rm r}^{*}$, given by
\begin{equation}
    P_{\rm r}^{*}(x^{*},y^{*}) = \int_{0}^{L_z} \left( -\rho_\mathrm{f} \langle \hat{u}^{*} \hat{v}^{*} \rangle \frac{\partial \langle u^{*} \rangle}{\partial y^{*}}\right) dz^{*}.
\end{equation}
This quantity depicts the energy transfer rate between the mean flow and instability, or more generally, the rate of perturbation generation (turbulence generationin fully-developed turbulent flows~\cite{Davies_1997,Hussein_2015,Prandtl_1922,Morris_1976,Cossu_2004,Cimarelli_2019}).~Without control, the production rate is generally positive for an unstable laminar flow, indicating a flow resonance phenomenon where energy is being transferred from the mean flow to the instability, causing it to grow as it propagates downstream$-$hence the positive, upward trend of $K_\mathrm{p}^{*}$ that we observe in Figs.~\ref{Fig7}b and~\ref{Fig10}b.~In contrast, a negative production rate is a PSub-induced flow anti-resonance phenomenon whereby energy is transferred from the instability back to the mean flow.~A negative production rate of the perturbation kinetic energy diminishes the intensity of an instability. In Fig.~\ref{Fig15}, we present the production rate of perturbation kinetic energy (expressed in dimensionless form) with respect to the wall-normal direction at three streamwise $x$-locations (stations) for both strong (Fig.~\ref{Fig15}a) and weak (Fig.~\ref{Fig15}b) passive control.~In both plots, the nominal MM-based PSub with five unit cells is used. Since the TS waves are small linear perturbations, we observe significantly modest changes in the production rate, on the order of $\sim10^{-6}$ in dimensionless units; however, these changes elucidate the underlying dynamics of the impact of the PSub on the flow field.~In Fig.~\ref{Fig15}, Station 1 is at the left edge of the PSub, $x^*/\delta=6.25$ (solid curves). This is the position where the flow first ``experiences" the influence of the PSub, and according to Fig.~\ref{Fig10}b, where the strongest reduction in $K_\mathrm{p}^{*}$ occurs for the stabilization cases.~Station 2 is at the right edge of the PSub, $x^*/\delta=7.86$ (dashed-dotted curves). This is the location where the instability initiates its recovery from the effect of the PSub.~For the stabilization cases, at this station, we notice the perturbation kinetic energy rises substantially, exceeding even the rigid-wall case, but only for a short distance downstream.~The last streamwise station, Station 3, is at $x^*/\delta=12$ (dashed curves) which is at the far downstream where the effect of the PSub has practically vanished,~confirming that the influence of the PSub is strictly local, within and very closely around the control region.~Similar but opposite trends for $P_{\rm r}^{*}$ are observed for the destabilization cases.~A comparison between Figs.~\ref{Fig15}a and~\ref{Fig15}b clearly reveals that the absolute strength of the production rate of perturbation kinetic energy is larger for strong PSub control.~This is again consistent with the prediction of the performance metric $P$ from Fig.~\ref{Fig10}a.~The impact of the PSub on production rate along the $y$-direction is also intriguing, showing that it starts with zero at the wall (due to the nominally zero velocity boundary conditions), reaches the peak close to the wall, and then gradually diminishes to zero again around the centerline. The near wall peak of the production rate occurs closer to the wall.~Moreover, due to the existence of the PSub at the bottom wall, the flow is not symmetric along wall-normal direction, see Fig.~\ref{FigAppA}. ~An
analysis of the flux of the perturbation energy for PnC-based PSubs is provided in Ref.~\cite{Hussein_2015}.
%\begin{figure*} [h!]
%	\includegraphics[width=1\columnwidth]{Figure_06.eps}%
%	\caption{Four key characterization plots for MM-based PSub: (a) Dispersion curves for a unit cell from which MM-based PSub is formed. Steady-state vibration (b) amplitude and (c) phase response at top edge of 10 unit-cell long PSub. (e) Performance metric obtained by multiplying the amplitude by the phase. The phase is between the force and the displacement at the PSub top edge. All plots are obtained by analysing a stand-alone FE model of the PSub without yet coupling to the flow. The results for a corresponding homogeneous structure are shown for comparison.} \label{Fig6}
%\end{figure*}

\section{Conclusions}
The theory of phononic subsurfaces enables the design of subsurface structures for the passive responsive control of wall-bounded laminar/transitional flows with growing instabilities.~We have investigated an MM-based configuration of PSubs that operates in the elastic subwavelength regime. This renders a PSub much shorter (5 cm) than the PnC-based PSub investigated in Ref.~\cite{Hussein_2015} (4 m).~We considered channel flows with unstable TS waves as examples for demonstrating the underlying performance of this new form of PSubs. A parallel analysis of a PnC-based PSub was conducted as well for comparison.~A PnC-based PSub is designed by tuning a stop-band truncation resonance to engage the target TS wave~\cite{Hussein_2015,Barnes_2021}, whereas the proposed MM-based PSub uses a pass-band resonance that has been lowered in frequency due to the generation of a subwavelength locally resonant band gap. 

Both TS wave stabilization and destabilization were demonstrated.~It was reaffirmed that the performance metric curve $P$ for a given PSub design (which is calculated \textit{a priori} without the need for coupled fluid-structure simulations) perfectly predicts both the nature of engagement with the instability (i.e., stabilization versus destabilization) and the intensity of engagement (e.g., weak, moderate, or strong control of the instability).~The results clearly display that the perturbation kinetic energy of the flow instability field is altered as desired specifically near the wall in the channel region where the PSub is installed.~Furthermore, and importantly, it was shown that the time-averaged value of $K_\mathrm{p}^{*}$ returns to nearly the same level as the reference rigid-wall case downstream of the PSub.~This ascertains the \textit{local} nature of PSub-based flow control, which in turn implies the ability to extend control to wider spatial regions by installing more PSubs as desired.~The time-averaged total elastodynamic energy in the PSub was also calculated and shown to be relatively low, zero, or high for stabilization, neutral effect, or destabilization, respectively.~This demonstrates the coherent nature of the PSub controlled coupled fluid-structure interaction and phased response across both media, and confirms the perfect predictability of the actual response by the predetermined value of the performance metric $P$.~Analysis of the rate of production of the flow perturbation kinetic energy, as a function of both the downstream and wall-normal directions, reveals the intrinsic anti-resonance and resonance mechanisms that take place within the flow when a PSub is installed. For stabilization, a PSub causes steady-state energy transfer from the flow instability into the mean flow at the start of the control region and vice versa closer to its end.~The opposite effect takes place for a PSub designed to destabilize the flow.

The PSubs theory lays the foundation for a mechanistic, spatially precise, and frequency- and wavenumber-dependent passive and responsive flow control paradigm that is fundamentally based on enabling a targeted contiguous synchronization of wave characteristics across both the flow and an interfacing subsurface elastic structure.~Future research will aim to advance PSubs design to enlarge the green area ($A_{\rm P}^{\rm S}$) or red area ($A_{\rm P}^{\rm D}$) under the $P$ curve in Fig.~\ref{Fig10}a for flow stabilization or destabilization, respectively. Emphasis will be on both deepening and widening these green and red regions to further strengthen the control and make it more robust over broad-frequency ranges. Ongoing innovative research in phononics (see reviews by Hussein et al.~\cite{Hussein_2014}, Jin et al.~\cite{Jin_2021}, and others) will drive this track.~Investigation of PSubs will be extended to boundary-layer flows, supersonic and hypersonic flows, advanced transitional flows, and fully developed turbulent flows, among other problems in flow control~\cite{Gad-el-Hak_2000}.~Switchable PSub control using piezoelectrics~\cite{Hagood_1991,Thorp_2001} is also a potential application.~Multifunctional PSub design to target flow control and, simultaneously, vibroacoustic control~\cite{Bilal_2018}, energy harvesting~\cite{Patrick_2021}, and/or structural support~\cite{Meza_2015} is another promising research direction that will build on the current investigation.

\section*{Acknowledgement}
The authors dedicate this paper to the memory of Professor Sedat Biringen (1945-2020). This work utilized the RMACC Summit supercomputer, which is supported by the National Science Foundation (awards ACI-1532235 and ACI-1532236),
the University of Colorado Boulder, and Colorado State University. The Summit supercomputer is a joint effort of the University of Colorado Boulder and Colorado State University.

\bigskip

\bibliographystyle{ieeetr}
\bibliography{Refs}

\begin{thebibliography}{10}

\bibitem{Gad-el-Hak_2000}
M.~Gad-el Hak, {\em Flow control: passive, active, and reactive flow
  management}.
\newblock Cambridge University Press, Cambridge, 2000.

\bibitem{Wehrmann_1965}
O.~H. Wehrmann, ``Tollmien-schlichting waves under the influence of a flexible
  wall,'' {\em Physics of Fluids}, vol.~8, pp.~1389--1390, 1965.

\bibitem{Liepmann_1982}
H.~W. Liepmann and D.~N. Nosenchuck, ``Active control of laminar-turbulent
  transition,'' {\em Journal of Fluid Mechanics}, vol.~118, pp.~201--204, 1982.

\bibitem{Joslin_1995}
R.~D. Joslin, R.~A. Nicolaides, G.~Erlebacher, M.~Y. Hussaini, and M.~D.
  Gunzburger, ``Active control of boundary-layer instabilities: Use of sensors
  and spectral controller,'' {\em AIAA Journal}, vol.~33, pp.~1521--1523, 1995.

\bibitem{Grundmann_2008}
S.~Grundmann and C.~Tropea, ``Active cancellation of artificially introduced
  tollmien–schlichting waves using plasma actuators,'' {\em Experiments in
  Fluids}, vol.~44, pp.~795--806, 2008.

\bibitem{Amitay_2016}
M.~Amitay, B.~A. Tuna, and H.~Dell’Orso, ``Identification and mitigation of
  {T-S} waves using localized dynamic surface modification,'' {\em Physics of
  Fluids}, vol.~28, p.~064103, 2016.

\bibitem{Jansen_2018}
K.~Jansen, M.~Rasquin, J.~Farnsworth, N.~Rathay, M.~Monastero, and M.~Amitay,
  ``Interaction of a synthetic jet with separated flow over a vertical tail,''
  {\em AIAA Journal}, vol.~56, pp.~2653--2668, 2018.

\bibitem{Walsh_1978}
M.~J. Walsh and L.~M. Weinstein, ``Drag and heat transfer on surfaces with
  small longitudinal fins,'' in {\em 11th Fluid and Plasma Dynamics Conference,
  Seatle, Washington, USA, July 11-12}, 1978.

\bibitem{Garcia_2011}
R.~Garc\'{i}a-Mayoral and J.~Jim\'{e}nez, ``Drag reduction by riblets,'' {\em
  Philosophical Transactions of the Royal Society A}, vol.~369, pp.~1412--1427,
  2011.

\bibitem{Cossu_2002}
C.~Cossu and L.~Brandt, ``Stabilization of {T}ollmien–{S}chlichting waves by
  finite amplitude optimal streaks in the blasius boundary layer,'' {\em
  Physics of Fluids}, vol.~14, pp.~L57--L60, 2002.

\bibitem{Fransson_2005}
J.~H.~M. Fransson, L.~Brandt, A.~Talamelli, and C.~Cossu, ``Experimental study
  of the stabilization of {T}ollmien–{S}chlichting waves by finite amplitude
  streaks,'' {\em Physics of Fluids}, vol.~17, p.~054110, 2005.

\bibitem{Abderrahaman_2017}
N.~Abderrahaman-Elena and R.~García-Mayoral, ``Analysis of anisotropically
  permeable surfaces for turbulent drag reduction,'' {\em Physical Review
  Fluids}, vol.~2, p.~114609, 2017.

\bibitem{Kramer_1957}
M.~O. Kramer, ``Boundary layer stabilization by distributed damping,'' {\em
  Naval Engineers Journal}, vol.~74, no.~2, pp.~341--348, 1962.

\bibitem{Benjamin_1960}
T.~B. Benjamin, ``Effects of a flexible boundary on hydrodynamic instability,''
  {\em Journal of Fluid Mechanics}, vol.~9, pp.~513--532, 1960.

\bibitem{Bushnell_1977}
D.~M. {Bushnell}, J.~N. {Hefner}, and R.~L. {Ash}, ``{Effect of compliant wall
  motion on turbulent boundary layers},'' {\em Physics of Fluids}, vol.~20,
  pp.~S31--S48, Oct. 1977.

\bibitem{Gad-el-Hak_1984}
M.~Gad-El-Hak, R.~F. Blackwelder, and J.~J. Riley, ``On the interaction of
  compliant coatings with boundary-layer flows,'' {\em Journal of Fluid
  Mechanics}, vol.~140, p.~257–280, 1984.

\bibitem{Carpenter_1985}
P.~W. Carpenter and A.~D. Garrad, ``The hydrodynamic stability of flow over
  {K}ramer-type compliant surfaces. part 1. tollmien-schlichting
  instabilities,'' {\em Journal of Fluid Mechanics}, vol.~155, p.~465–510,
  1985.

\bibitem{Lucy_1995}
A.~D. Lucey and P.~W. Carpenter, ``Boundary layer instability over compliant
  walls: {C}omparison between theory and experiment,'' {\em Physics of Fluids},
  vol.~7, pp.~2355--2363, 1995.

\bibitem{Davies_1997}
C.~Davies and P.~W. Carpenter, ``Numerical simulation of the evolution of
  tollmien-schlichting waves over finite compliant panels,'' {\em Journal of
  Fluid Mechanics}, vol.~335, pp.~361--392, 1997.

\bibitem{Luhar_2015}
M.~Luhar, A.~S. Sharma, and B.~J. McKeon, ``A framework for studying the effect
  of compliant surfaces on wall turbulence,'' {\em Journal of Fluid Mechanics},
  vol.~768, pp.~415--441, 2015.

\bibitem{Esteghamatian_2022}
A.~Esteghamatian, J.~Katz, and T.~A. Zaki, ``Spatiotemporal characterization of
  turbulent channel flow with a hyperelastic compliant wall,'' {\em Journal of
  Fluid Mechanics}, vol.~942, p.~A35, 2022.

\bibitem{Hussein_2015}
M.~I. Hussein, S.~Biringen, O.~R. Bilal, and A.~Kucala, ``Flow stabilization by
  subsurface phonons,'' {\em Proceedings of the Royal Society A}, vol.~471,
  p.~20140928, 2015.

\bibitem{Barnes_2021}
C.~J. Barnes, C.~L. Willey, K.~Rosenberg, A.~Medina, and A.~T. Juhl, {\em
  Initial Computational Investigation Toward Passive Transition Delay Using a
  Phononic Subsurface}.

\bibitem{Tollmien_1928}
W.~Tollmien, ``Über die entstehung der turbulenz. 1. mitteilung,'' {\em
  Nachrichten von der Gesellschaft der Wissenschaften zu Göttingen,
  Mathematisch-Physikalische Klasse}, vol.~1929, pp.~21--44, 1928.

\bibitem{Schlichting_1933}
H.~Schlichting, ``Zur enstehung der turbulenz bei der plattenströmung,'' {\em
  Nachrichten von der Gesellschaft der Wissenschaften zu Göttingen,
  Mathematisch-Physikalische Klasse}, vol.~1933, pp.~181--208, 1933.

\bibitem{Hussein_2014}
M.~I. Hussein, M.~J. Leamy, and M.~Ruzzene, ``{Dynamics of Phononic Materials
  and Structures: Historical Origins, Recent Progress, and Future Outlook},''
  {\em Applied Mechanics Reviews}, vol.~66, 05 2014.
\newblock 040802.

\bibitem{Jin_2021}
Y.~Jin, Y.~Pennec, B.~Bonello, H.~Honarvar, L.~Dobrzynski, B.~Djafari-Rouhani,
  and M.~I. Hussein, ``Physics of surface vibrational resonances: pillared
  phononic crystals, metamaterials, and metasurfaces,'' {\em Reports on
  Progress in Physics}, vol.~84, p.~086502, aug 2021.

\bibitem{wallis1957effect}
R.~F. Wallis, ``Effect of free ends on the vibration frequencies of
  one-dimensional lattices,'' {\em Phys. Rev.}, vol.~105, pp.~540--545, Jan
  1957.

\bibitem{Camley_1983}
R.~E. Camley, B.~Djafari-Rouhani, L.~Dobrzynski, and A.~A. Maradudin,
  ``Transverse elastic waves in periodically layered infinite and semiinfinite
  media,'' {\em Physical Review B}, vol.~27, pp.~7318--7329, 1983.

\bibitem{Davis_2011}
B.~Davis, A.~Tomchek, E.~Flores, L.~Liu, and M.~Hussein, ``Analysis of
  periodicity termination in phononic crystals,'' vol.~8, 01 2011.

\bibitem{Albabaa_2017}
H.~Al~Ba’ba’a, M.~Nouh, and T.~Singh, ``Pole distribution in finite
  phononic crystals: Understanding {B}ragg-effects through closed-form
  system,'' {\em Journal of the Acoustical Society of America}, vol.~142,
  pp.~1399--1412, 2017.

\bibitem{Bastawrous_2022}
M.~V. Bastawrous and M.~I. Hussein, ``Closed-form existence conditions for
  bandgap resonances in a finite periodic chain under general boundary
  conditions,'' {\em Journal of the Acoustical Society of America}, vol.~151,
  pp.~286--298, 2022.

\bibitem{Albabaa_2022}
H.~Al~Ba’ba’a, C.~L. Willey, V.~W. Chen, A.~T. Juhl, and M.~Nouh, ``Theory
  of truncation resonances in continuum rod-based phononic crystals with
  generally asymmetric unit cells,'' {\em arXiv:2211.01423v1}, 2022.

\bibitem{Rosa_2022}
M.~I.~N. Rosa, B.~L. Davis, L.~Liu, M.~Ruzzene, and M.~I. Hussein, ``Material
  vs. structure: Topological origins of band-gap truncation resonances in
  periodic structures,'' {\em arXiv:submit/4672353}, 2022.

\bibitem{Kushwaha_1993}
M.~S. Kushwaha, P.~Halevi, L.~Dobrzynski, and B.~Djafari-Rouhani, ``Acoustic
  band structure of periodic elastic composites,'' {\em Physical Review
  Letters}, vol.~71, pp.~2022--2025, Sep 1993.

\bibitem{Liu_2000}
Z.~Liu, X.~Zhang, Y.~Mao, Y.~Y. Zhu, Z.~Yang, C.~T. Chan, and P.~Sheng,
  ``Locally resonant sonic materials,'' {\em Science}, vol.~289, no.~5485,
  pp.~1734--1736, 2000.

\bibitem{Willey_2022}
C.~L. Willey, V.~W. Chen, D.~Roca, A.~Kianfar, M.~I. Hussein, and A.~T. Juhl,
  ``Coiled phononic crystal with periodic rotational locking: {S}ubwavelength
  {B}ragg band gaps,'' {\em Physical Review Applied}, vol.~18, p.~014035, 2022.

\bibitem{Deymier_2013}
P.~A. Deymier, ``Introduction to phononic crystals and acoustic
  metamaterials,'' in {\em Acoustic metamaterials and phononic crystals},
  pp.~1--12, Springer, Berlin, 2013.

\bibitem{Craster_2013}
R.~Craster and S.~Guenneau, {\em Acoustic Metamaterials: Negative Refraction,
  Imaging, Lensing and Cloaking}.
\newblock Springer, Dordrecht, 01 2013.

\bibitem{Phani_2017}
A.~S. Phani and M.~I. Hussein, {\em Introduction to Lattice Materials}, ch.~1,
  pp.~1--17.
\newblock John Wiley \& Sons, Ltd, 2017.

\bibitem{xiao2013flexural}
Y.~Xiao, J.~Wen, D.~Yu, and X.~Wen, ``Flexural wave propagation in beams with
  periodically attached vibration absorbers: band-gap behavior and band
  formation mechanisms,'' {\em J. Sound Vib.}, vol.~332, no.~4, pp.~867--893,
  2013.

\bibitem{Sangiuliano_2020}
L.~Sangiuliano, C.~Claeys, E.~Deckers, and W.~Desmet, ``Influence of boundary
  conditions on the stop band effect in finite locally resonant metamaterial
  beams,'' {\em Journal of Sound Vibration}, vol.~473, p.~115225, 2020.

\bibitem{Xia_2020}
Y.~Xia, A.~Erturk, and M.~Ruzzene, ``Topological edge states in quasiperiodic
  locally resonant metastructures,'' {\em Physical Review Applied}, vol.~13,
  p.~014023, 2020.

\bibitem{Park_2022}
S.~Park, G.~K. Hristov, S.~Balasubramanian, A.~G. Goza, P.~J. Ansell, and K.~H.
  Matlack, ``Design and analysis of phononic material for passive flow
  control,'' in {\em AIAA AvIATION FORUM, Chicago, Illinois, USA, June 27-July
  1}, 2022.

\bibitem{Bloch_1929}
F.~Bloch, ``{\"U}ber die quantenmechanik der elektronen in kristallgittern,''
  {\em Zeitschrift f{\"u}r physik}, vol.~52, no.~7-8, pp.~555--600, 1929.

\bibitem{Hussein_2009}
M.~I. Hussein, ``Reduced {B}loch mode expansion for periodic media band
  structure calculations,'' {\em Proceedings of the Royal Society A}, vol.~465,
  pp.~2825--2848, 2009.

\bibitem{HusseinJSV06}
M.~I. Hussein, G.~M. Hulbert, and R.~A. Scott, ``Dispersive elastodynamics of
  1{D} banded materials and structures: analysis,'' {\em Journal of Sound and
  Vibration}, vol.~289, no.~4, pp.~779--806, 2006.

\bibitem{Hussein_PRB_2009}
M.~I. Hussein, ``Theory of damped {B}loch waves in elastic media,'' {\em
  Physical Review B}, vol.~80, p.~212301, Dec 2009.

\bibitem{navier1823memoire}
C.~Navier, ``M{\'e}moire sur les lois du mouvement des fluides,'' {\em
  \textit{M{\'e}moires de l’Acad{\'e}mie Royale des Sciences de l’Institut
  de France}}, pp.~389--440, 1823.

\bibitem{stokes1845g}
G.~G. Stokes, ``{On the Theories of the Internal Friction of Fluids in Motion,
  and of the Equilibrium and Motion of Elastic Solids},'' in {\em {Classics of
  Elastic Wave Theory}}, Society of Exploration Geophysicists, 01 2007.

\bibitem{Orr_I_1907}
W.~M. Orr, ``The stability or instability of the steady motions of a perfect
  liquid and of a viscous liquid. part i: A perfect liquid,'' {\em Proceedings
  of the Royal Irish Academy. Section A: Mathematical and Physical Sciences},
  vol.~27, pp.~9--68, 1907.

\bibitem{Orr_II_1907}
W.~M. Orr, ``The stability or instability of the steady motions of a perfect
  liquid and of a viscous liquid. part ii: A viscous liquid,'' {\em Proceedings
  of the Royal Irish Academy. Section A: Mathematical and Physical Sciences},
  vol.~27, pp.~69--138, 1907.

\bibitem{Sommerfeld_1908}
A.~Sommerfield, ``Ein beitrag zur hydrodynamischen erklarung der turbulenten
  flussigkeisbewegung,'' in {\em Atti del IV Congresso internazionale dei
  Matematici}, 1908.

\bibitem{Nishioka_JFM_1975}
M.~Nishioka, S.~I. A, and Y.~Ichikawa, ``An experimental investigation of the
  stability of plane poiseuille flow,'' {\em Journal of Fluid Mechanics},
  vol.~72, no.~4, p.~731–751, 1975.

\bibitem{Schubauer_1948}
G.~B. Schubauer and H.~K. Skramstad, ``Laminar-boundary-layer oscillations and
  transition on a flat plate,'' {\em Journal of research of the National Bureau
  of Standards}, vol.~38, p.~251, 1947.

\bibitem{Klebanoff_JFM_1962}
P.~S. Klebanoff, K.~D. Tidstrom, and L.~M. Sargent, ``The three-dimensional
  nature of boundary-layer instability,'' {\em Journal of Fluid Mechanics},
  vol.~12, no.~1, p.~1–34, 1962.

\bibitem{Dana91}
G.~Danabasoglu, S.~Biringen, and C.~L. Streett, ``Spatial simulation of
  instability control by periodic suction blowing,'' {\em Physics of Fluids A:
  Fluid Dynamics}, vol.~3, no.~9, pp.~2138--2147, 1991.

\bibitem{Saiki93}
E.~M. Saiki, S.~Biringen, G.~Danabasoglu, and C.~L. Streett, ``Spatial
  simulation of secondary instability in plane channel flow: comparison of {K}-
  and {H}-type disturbances,'' {\em Journal of Fluid Mechanics}, vol.~253,
  p.~485–507, 1993.

\bibitem{Kucala14}
A.~Kucala and S.~Biringen, ``Spatial simulation of channel flow instability and
  control,'' {\em Journal of Fluid Mechanics}, vol.~738, p.~105–123, 2014.

\bibitem{Reynolds69}
W.~Reynolds, {\em Orrsom: a Fortran-IV Program for Solution of the
  Orr-Somerfield Equation}.
\newblock 1969.

\bibitem{Lighthill_1958}
M.~J. Lighthill, ``On displacement thickness,'' {\em Journal of Fluid
  Mechanics}, vol.~4, pp.~383--392, 1958.

\bibitem{Sankar_1981}
N.~L. Sankar, J.~B. Malone, and Y.~Tassa, ``An implicit conservative algorithm
  for steady and unsteady three-dimensional transonic potential flows,'' in
  {\em AIAA Paper 81-1016, June 1981}, 1981.

\bibitem{Farhat_2000}
C.~Farhat and M.~Lesoinne, ``Two effcient staggered algorithms for the serial
  and parallel solution of three-dimensional nonlinear transient aeroelastic
  problems,'' {\em Computer Methods in Applied Mechanics and Engineering},
  vol.~182, pp.~499--515, 2000.

\bibitem{wu2008evidence}
T.-T. Wu, Z.-G. Huang, T.-C. Tsai, and T.-C. Wu, ``Evidence of complete band
  gap and resonances in a plate with periodic stubbed surface,'' {\em Applied
  Physics Letters}, vol.~93, no.~11, p.~111902, 2008.

\bibitem{pennec2008low}
Y.~Pennec, B.~Djafari-Rouhani, H.~Larabi, J.~O. Vasseur, and A.-C.
  Hladky-Hennion, ``Low-frequency gaps in a phononic crystal constituted of
  cylindrical dots deposited on a thin homogeneous plate,'' {\em Physical
  Review B}, vol.~78, no.~10, p.~104105, 2008.

\bibitem{Bilal_2013}
O.~R. Bilal and M.~I. Hussein, ``Trampoline metamaterial: Local resonance
  enhancement by springboards,'' {\em Applied Physics Letters}, vol.~103,
  no.~11, p.~111901, 2013.

\bibitem{Xiao_2013}
Y.~Xiao, J.~Wen, D.~Yu, and X.~Wen, ``Flexural wave propagation in beams with
  periodically attached vibration absorbers: Band-gap behavior and band
  formation mechanisms,'' {\em Journal of Sound and Vibration}, vol.~332,
  no.~4, pp.~867--893, 2013.

\bibitem{Khajehtourian_2014}
R.~Khajehtourian and M.~I. Hussein, ``Dispersion characteristics of a nonlinear
  elastic metamaterial,'' {\em AIP Advances}, vol.~4, no.~12, p.~124308, 2014.

\bibitem{Prandtl_1922}
L.~{Prandtl}, ``{Bemerkungen {\"u}ber die Entstehung der Turbulenz},'' {\em
  Zeitschrift Angewandte Mathematik und Mechanik}, vol.~1, pp.~431--436, Jan.
  1921.

\bibitem{Morris_1976}
P.~J. Morris, ``The spatial viscous instability of axisymmetric jets,'' {\em
  Journal of Fluid Mechanics}, vol.~77, pp.~511--529, 1976.

\bibitem{Cossu_2004}
C.~Cossu and L.~Brandt, ``On {T}ollmien–{S}chlichting-like waves in streaky
  boundary layers,'' {\em European Journal of Mechanics B/Fluids}, vol.~23,
  pp.~815--833, 2004.

\bibitem{Cimarelli_2019}
A.~Cimarelli, A.~Leonforte, E.~{De Angelis}, A.~Crivellini, and D.~Angeli, ``On
  negative turbulence production phenomena in the shear layer of separating and
  reattaching flows,'' {\em Physics Letters A}, vol.~383, no.~10,
  pp.~1019--1026, 2019.

\bibitem{Hagood_1991}
N.~W. Hagood and A.~Von~Flotow, ``Damping of structural vibrations with
  piezoelectric materials and passive electrical networks,'' {\em Journal of
  Sound and Vibration}, vol.~146, pp.~243--268, 1991.

\bibitem{Thorp_2001}
O.~Thorp, M.~Ruzzene, and A.~Baz, ``Attenuation and localization of wave
  propagation in rods with periodic shunted piezoelectric patches,'' {\em Smart
  Materials and Structures}, vol.~10, pp.~979--989, 2001.

\bibitem{Bilal_2018}
O.~R. Bilal, D.~Ballagi, and C.~Daraio, ``Architected lattices for simultaneous
  broadband attenuation of airborne sound and mechanical vibrations in all
  directions,'' {\em Physical Review Applied}, vol.~10, p.~054060, 2018.

\bibitem{Patrick_2021}
J.~Patrick, S.~Adhikari, and M.~I. Hussein, ``Brillouin-zone characterization
  of piezoelectric material intrinsic energy-harvesting availability,'' {\em
  Smart Materials and Structures}, vol.~30, p.~085022, jul 2021.

\bibitem{Meza_2015}
L.~R. Meza, A.~J. Zelhofer, N.~Clarke, A.~J. Mateos, D.~M. Kochmann, and J.~R.
  Greer, ``Resilient 3d hierarchical architected metamaterials,'' {\em
  Proceedings of the National Academy of Sciences}, vol.~112, pp.~11502--11507,
  2015.

\end{thebibliography}

\appendix

\section{Production of the perturbation kinetic energy across entire channel cross-section}
\label{sec:appa}
\begin{figure*} [b!]
 	\includegraphics[width=1\columnwidth]{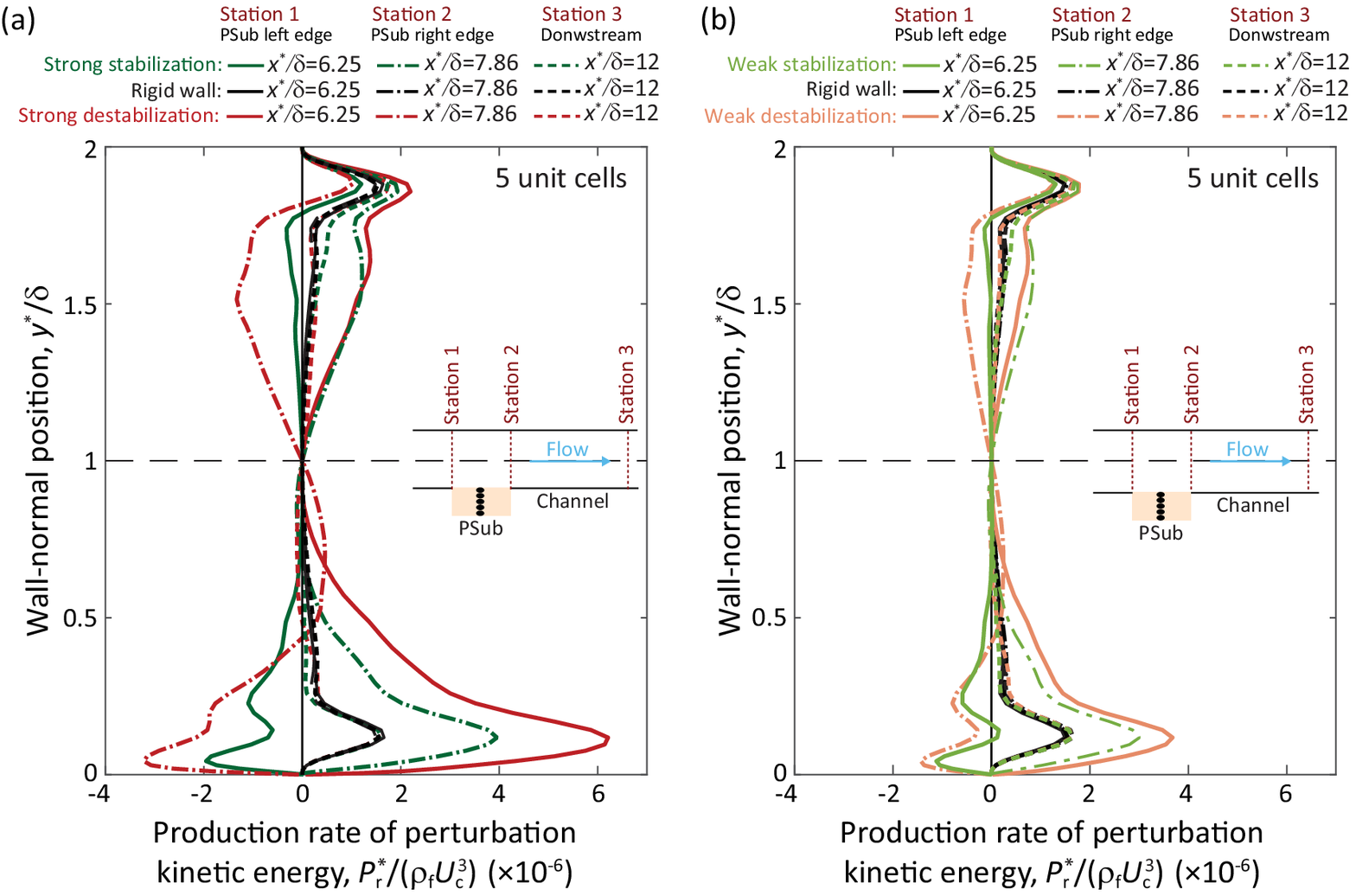}%
	\caption{Production of flow perturbation kinetic energy for the cases considered in Fig.~\ref{Fig15}, but plotted across the entire channel height.} \label{FigAppA}
\end{figure*}

In Fig.~\ref{Fig15}, we show the rate of production of perturbation kinetic energy as a function of the wall-normal direction over the bottom half of the channel where the PSub is applied. In Fig.~\ref{FigAppA}, we show the same result but extend the view to cover the entire channel height.~We observe a zero rate of production at the channel center line and some PSub effects, although much weakened at the top half of the channel. 

\section{Production of the perturbation (instability) kinetic energy for PSubs installed at bottom and top of channel walls.}
\label{sec:appb}

\begin{figure*} [t!]
	\includegraphics[width=1\columnwidth]{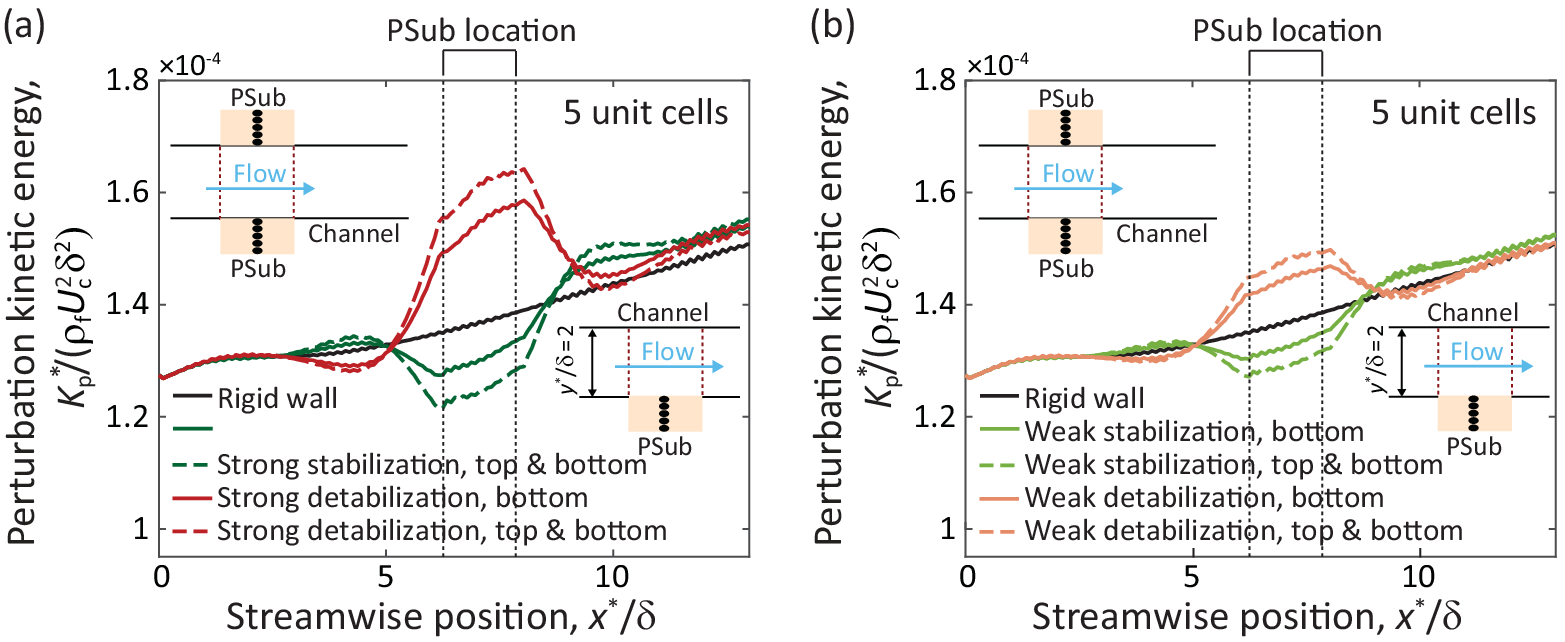}%
	\caption{~Time-averaged kinetic energy of the perturbation (instability) of the (a) strongest and (b) weakest stabilization and destabilization cases when two PSubs are installed, one at the bottom wall and the other at the top wall.~The PSub location spans the distance between the two dashed lines as indicated.~The presence of two PSubs increases the intensity of the stabilization or destabilization.} \label{FigAppB1}
\end{figure*}

\begin{figure*} [b!]
	\includegraphics[width=1\columnwidth]{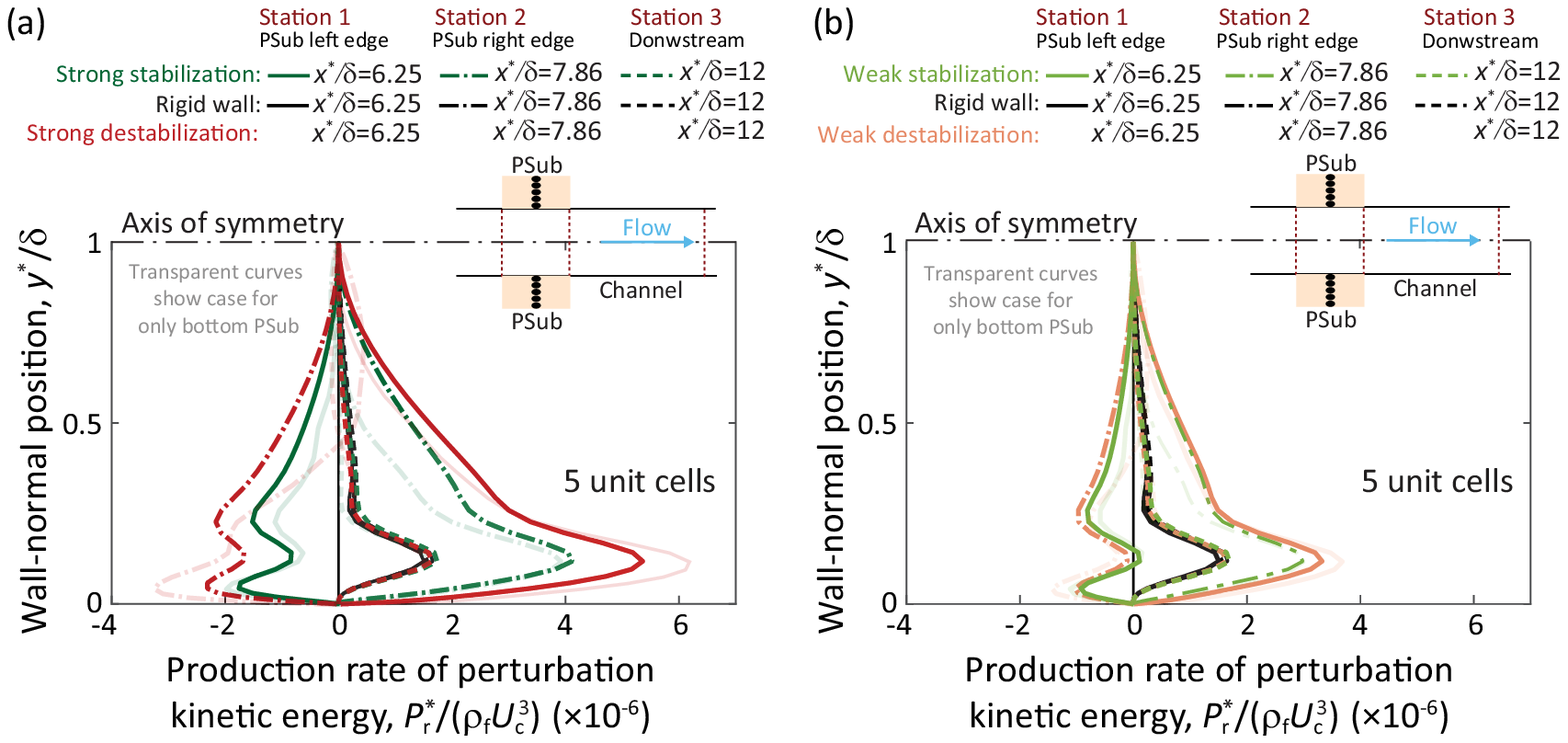}%
	\caption{Production of flow perturbation (instability) kinetic energy for the cases considered in Fig.~\ref{FigAppB1}, but showing only the bottom half of the channel due to symmetry.~Results for only bottom PSub are shown as transparent curves for direct comparison.} \label{FigAppB2}
\end{figure*}

In Fig.~\ref{FigAppB1}, we show results obtained when two PSubs are installed, one at the bottom (as in all the previous cases) and an additional one placed at the top. Thus in the absence of any other disturbances or stochastic variations, the flow remains symmetric in the wall-normal direction $y=1$ around the centerline axis of symmetry.~Clearly the intensity of the time-averaged kinetic energy changes over the entire channel section increases when two PSubs are applied.~This is shown for both the strong (Fig.~\ref{FigAppB1}a) and weak (Fig.~\ref{FigAppB1}b) stabilization and destabilization cases.~The corresponding results for the rate of production of perturbation kinetic energy are shown in Fig.~\ref{FigAppB2}.~These results indicate promise for the future application of PSubs around the entire circumference of long-range pipelines. 

\end{document}